\documentstyle[10pt]{report}

\begin{document}

\title{Quantum Noise in Quantum Optics:\\
       the Stochastic Schr\"odinger Equation}

\author{Peter Zoller ${}^\dagger$ and C.~W.~Gardiner ${}^*$\\[3mm]
${}^\dagger$ Institute for Theoretical Physics, 
University of Innsbruck,\\
6020 Innsbruck,
Austria \\[1mm]
${}^*$ Department of Physics, Victoria University\\
Wellington, New Zealand}

\maketitle


\input psfig.sty
\newcommand{\complaint}[1]{\vskip 5mm\hrule\vskip3mm #1 \vskip2mm\hrule
\vskip5mm}

\begin{abstract}
Lecture Notes for the Les Houches Summer School LXIII on Quantum Fluctuations
in July 1995:
``{\it Quantum Noise in Quantum Optics: the Stochastic Schroedinger Equation}''
to appear in Elsevier Science Publishers B.V. 1997,  
edited by E. Giacobino and S. Reynaud.
\end{abstract}

\newcommand{\debug}[1]{
}

\renewcommand{\le}{\leq} \renewcommand{\ge}{\geq}

\newcommand{\comment}[1]{}

\newcommand{\bea}{\begin{eqnarray}} \newcommand{\eea}{\end{eqnarray}}
\newcommand{\beann}{\begin{eqnarray*}} \newcommand{\eeann}{\end{eqnarray*}}
\newcommand{\be}{\begin{equation}}\newcommand{\ee}{\end{equation}}
\newcommand{\benn}{\[}\newcommand{\eenn}{\]}
\newcommand{\Heff}{H_{\rm eff}}
\newcommand{\half}{\frac{1}{2}}
\newcommand{\Sc}{ {\cal S}} 
\newcommand{\Tc}{ {\cal T}}
\newcommand{\Jc}{ {\cal J}}
\newcommand{\Lc}{ {\cal L}}
\newcommand{\Tr}{ {\rm Tr} }
\newcommand{\Ueff}{ U _{\rm eff} }

\newcommand{\ket}[1]{|#1  \rangle}
\newcommand{\bra}[1]{\langle #1  |}
\newcommand{\ME}[2]{\langle #1  |#2  \rangle }

\newcommand{\pct}{{\tilde \psi}_c}
\newcommand{\norm}[1]{ |\!| #1 |\!|}
\newcommand{\abs}[1]{ | #1 |}
\renewcommand{\vec}[1]{\stackrel{\rightarrow}{#1}}

\newcommand{\mycase}[4]{ %
\left\{ \begin{array}{ll} #1 & #2 %
\\  #3 & #4 \end{array} \right.%
}

\newcommand{\g}[1]{
}
\newcommand{\gs}[1]{
}

\newcommand{\AV}[1]{ \left\langle #1  \right\rangle}
\newcommand{\AVS}[1]{ \AV{ #1}_{\rm st}}
\newcommand{\AVc}[1]{ \AV{ #1}_{c}}
\newcommand{\Ito}{Ito}

\newcommand{\nn}{\nonumber}

\newcommand{\one}{\hat 1}
\newcommand{\OP}[1]{{\hat #1}}

\newcommand{\dt}{\; dt}

\newcommand{\nofrac}[2]{ #1 / #2  }

\newcommand{\cop}[1]{ {c}_{#1}}

\newcommand{\targ}[1]{ (#1)}
\newcommand{\OPU}{\OP{U}}
\newcommand{\OPV}{\OP{V}}
\newcommand{\out}{{\rm out}}
\newcommand{\inp}{{\rm in}}

\newcommand{\OPVout}{\OP{V}^\out}

\newcommand{\dB}[1]{dB\targ{#1}}
\newcommand{\dL}[1]{d\Lambda\targ{#1}}

\newcommand{\N}[2]{N_{#1}{(#2)}}
\newcommand{\dN}[2]{d\N{#1}{#2}}
\newcommand{\Y}[2]{Y_{#1}{(#2)}}
\newcommand{\dY}[2]{d\Y{#1}{#2}}

\newcommand{\Hsys}{H _{\rm sys}}
\newcommand{\HB}{H _{\rm B}}
\newcommand{\Hint}{H _{\rm int}}

\newcommand{\Ii}{} 
\newcommand{\Si}{\mbox{ \bf (S)}}

\renewcommand{\labelenumi}{(\roman{enumi})}


\newenvironment{Remark}{\begin{list}{}%
{
  \labelwidth0.0cm  \labelsep0.0cm
  \parsep0.5ex plus0.2ex minus 0.1ex \itemsep0ex plus 0.2ex
  \setlength{\topsep}{1.75\baselineskip}
  \setlength{\leftmargin}{\parindent}  
   \setlength{\rightmargin}{\leftmargin}    
   \renewcommand{\baselinestretch}{0.9}
\small \item[] }}%
{\end{list}}

\newenvironment{mylistnn}{\begin{list}{}%
{
  \labelwidth0.0cm \leftmargin0.5cm \labelsep0.0cm \rightmargin0.5cm
  \parsep0.5ex plus0.2ex minus 0.1ex \itemsep0ex plus 0.2ex \normalsize }}%
{\end{list}}

\newcounter{mylist}
\newenvironment{mylist}{\begin{list}{(\roman{mylist})}%
{\usecounter{mylist}
  \labelwidth3ex \leftmargin4ex \labelsep1ex \rightmargin4ex
  \parsep0.5ex plus0.1ex minus 0.1ex \itemsep0ex plus 0.2ex \normalsize }}%
{\end{list}}

%
\newcounter{Discussion}
\newenvironment{Discussion}{\begin{list}{(\roman{Discussion})}%
{\usecounter{Discussion}
  \labelwidth3ex \leftmargin4ex 
  \labelsep1ex 
  \rightmargin0ex 
  \setlength{\parsep}{\parskip} 
\topsep0ex
  \parsep0.5ex plus0.1ex minus 0.1ex \itemsep0ex plus 0.2ex \normalsize }}%
{\end{list}}

%
\newenvironment{ListNoNumber}{\begin{list}{}%
{
  \setlength{\leftmargin}{\parindent} 
  \labelsep1ex 
 \setlength{\rightmargin}{\leftmargin} 
  \setlength{\parsep}{\parskip} 
  \parsep0.5ex plus0.1ex minus 0.1ex \itemsep0ex plus 0.2ex \normalsize }}%
{\end{list}}

{\newlength{\egwidth}\setlength{\egwidth}{0.40\textwidth}
\newenvironment{eg}%
 {\begin{list}{}{\setlength{\leftmargin}{0.05\textwidth}%
 \setlength{\rightmargin}{\leftmargin}}\item[]\normalsize}%
{\end{list}}
\newenvironment{egbox}%
 {\begin{minipage}[t]{\egwidth}}%
 {\end{minipage}}
\newcommand{\egstart}{\begin{eg}\begin{egbox}}%
\newcommand{\egmid}{\end{egbox}\hfill\begin{egbox}}
\newcommand{\egend}{\end{egbox}\end{eg}}

\newcommand{\Cite}[1]{\cite{#1}}
\newcommand{\Ref}[1]{\ref{#1}\debug{\fbox{\sl\mbox{\tiny ref\{#1\}}}}}%
\newcommand{\Label}[1]{\label{#1}\debug{{\fbox{\tiny #1}}}}%
\newcommand{\Bibitem}[1]{\bibitem{#1}\debug{{\fbox{\tiny #1}}}}%

\newcommand{\Nd}{{N_c}}

\newcommand{\Title}[1]{}


\section{Introduction}

Theoretical quantum optics studies ``open systems,''  i.e. systems 
coupled to an ``environment'' \Cite{QN,Carmichael,WallsMilburn,Louisell}. In
quantum optics this environment corresponds to the infinitely many modes of the
electromagnetic field. The starting point of a description of quantum
noise is a modeling in terms of quantum Markov processes \Cite{QN}.
From a physical point of view, a quantum Markovian description is an
approximation where the environment is modeled as a heatbath with a
short correlation time and weakly coupled to the system.  In the
system dynamics the coupling to a bath will introduce {\em damping}
and {\em fluctuations}.  The radiative modes of the heatbath also
serve as {\em input channels} through which the system is driven, and
as {\em output channels} which allow the continuous observation of the
radiated fields.  Examples of quantum optical systems are resonance
fluorescence, where a radiatively damped atom is driven by laser light
(input) and the properties of the emitted fluorescence light (output)are
measured,
and an optical cavity mode coupled to the outside radiation modes by a
partially transmitting mirror.

 Historically, the first formulations were given in terms of quantum
Langevin equations (as developed in the context of laser theory) which
are Heisenberg equations for system operators with the reservoir
eliminated in favor of damping terms and white noise operator forces
(see references in \Cite{QN}).  The alternative and equivalent
formulation in terms of a master equation for a reduced system density
operator together with the quantum fluctuation regression theorem, has
provided the most important practical tool in quantum optics, in
particular for nonlinear systems \Cite{QN,Carmichael}.

The rigorous mathematical basis for these methods is quantum
stochastic calculus (QSC) as formulated by Hudson and Parthasarathy
\Cite{HP}.  QSC is a noncommutative analogue of Ito's stochastic
calculus \Cite{SM}. The basic ingredients are ``white noise''
Bose fields $b(t)$, $b(t)^\dagger$ with canonical commutation
relations $[b(t),b(t')^\dagger]=\delta (t-t')$. In quantum optics
these Bose fields can be considered as an approximation to the
electromagnetic field.  The connection between these abstract
mathematical developments and the physical principles and foundations
of quantum optics is discussed in particular in the work of Barchielli
\Cite{BL85,B86,B87,B90,BB91}, and Gardiner and coworkers 
\Cite{QN,GardinerCollett,GPZ92}. 
Gardiner and Collett \Cite{GardinerCollett} showed the connection
between the more physically motivated quantum Langevin equations and the
more mathematically precise ``quantum stochastic differential
equations'' (QSDE). Furthermore, these authors introduced as an
essential element of the theory an interpretation of the Bose--fields
as ``input'' and ``output'' fields corresponding to the field before
and after the interaction with the system. QSC allows also the
development of a consistent theory of photodetection (photon counting
and heterodyne measurements) \Cite{BB91,WisemanThesis} in direct
relationship with the theory of continuous measurement of Srinivas and
Davies
\Cite{SD81}.

While most of the theoretical work in quantum optics has emphasized
quantum Langevin and master equations \Cite{QN}, recent developments and
applications have focused on formulations employing the {\em quantum
stochastic Schr\"odinger equation} (QSSE) \Cite{B87,B90,GPZ92,G94}
 and its c-number
version, a {\em stochastic Schr\"odinger equation} (SSE)
\Cite{Carmichael,B90,BB91,GPZ92,%
WisemanThesis,G94,DCM92,M93,DRZ92,DPZG92,%
TC92,N95,Bel,CBA92,Marte,Dum94,CBSDM94,NKS95,CM95,SGK95,GK94a,GK94b,C93}. 
Phenomenological SSEs have been given in Refs.~\Cite{G84,GP92,Stenholm,DGHP95}.
The QSSE is a QSDE for the state vector of the combined system +
heatbath with Bose fields $b(t)$ and $b(t)^\dagger$ as noise
operators. This equation generates a unitary time evolution, and
implies the master equation and the quantum regression theorem for
multitime averages, and the Quantum Langevin equation for system
operators as exact results.

The SSE, on the other hand, is a stochastic evolution equation for the
system wave function with damping and c--number noise terms which
implies a non-unitary time evolution.  The relationship between the
QSSE and the SSE is established by the continuous measurement
formalism of Srinivas and Davies for counting processes (photon
counting)\Cite{SD81}.  This continuous measurement theory has the
interpretation as a probabilistic description in terms of quantum
jumps in a system. Thus the continuous measurement theory can be {\em
simulated} probabilistically, and this
simulation yields a sequence of system wave functions with jumps at
times $t_1,t_2,\ldots,t_n$ by a rule which is directly related to the
structure of the appropriate QSDE.  The trajectories of counts
generated in these simulations have the same statistics as the photon
statistics derived within the standard photodetection theory, and in
this sense the individual counting sequences generated in a single
computer run can, with some caution, be interpreted as ``what an
observation would yield in a single run of the experiment.''  These
simulations can thus {\em illustrate} the dynamics of single quantum
systems, for example in the quantum jumps in ion traps \Cite{ZMW87} or the
squeezing dynamics \Cite{Carmichael}. In continuous measurement
theory the time evolution of the system conditional on having observed
(or simulated) a certain count sequence is called {\it a posteriori
dynamics}. The master equation is recovered in this description as
ensemble average over all counting trajectories. Again, in the
language of continuous measurement theory, this corresponds to a
situation where the measurements (the counts) are not read, i.e. no
selection is made ({\it a priori dynamics}). Simulation of the SSE and
averaging over the noise provides an new computational tool to
generate solutions of the master equation. An important feature, first
emphasized by Dalibard, Castin and M{\o}lmer \Cite{DCM92} is that one
only has to deal with a wave function of dimension $N$, as opposed to
working with the density matrix which has $N^2$ elements. Thus,
simulations can provide solutions when a direct solution of the
master equation is impractical because of the large dimension of the
system space.

When instead of direct photon counting we perform a homodyne
experiment, where the system output is mixed with a local oscillator,
and a homodyne current is measured, the jumps are replaced by a
diffusive evolution.  These diffusive Schr\"odinger equations were
first derived by Carmichael \Cite{Carmichael,WisemanThesis} from his
analysis of homodyne detection, and independently in a more formal
context by Barchielli and Belavkin \Cite{BB91}. Somewhat earlier
equations of this form have been postulated in connection with
dynamical theories of wave function reduction \Cite{G84,GP92,DGHP95}.

In these notes we will review in a pedagogical way some of the recent
developments of quantum noise methods, continuous measurement and the
stochastic Schr\"odinger equation and some applications, and the main
purpose is to emphasize and illustrate the physical basis of such a
formulation.

\section{An introductory example: Mollow's pure state analysis of resonant
light scattering}

As an introduction and historical remark we briefly review Mollow's
work on the {\em pure state analysis of resonant light scattering} \Cite{M75}.
Twenty years ago Mollow developed a theory of resonance fluorescence of
a strongly driven two--level atom, where he showed that the atomic
density matrix could be decomposed into contributions from
subensembles corresponding to a certain trajectory of photon
emissions, which could be represented by {\em atomic wave functions}.
This formulation anticipated and provided the basis
\Cite{GPZ92,DRZ92,ZMW87} for some of the recent developments in
quantum optics from the Stochastic Schr\"odinger Equation for system
wave functions to our understanding of the photon statistics in the
framework of continuous measurement theory.  Below we will briefly review some
of Mollow's ideas in a more modern language, closely related
to the discussions of the following  sections.

\subsubsection{Hierarchy of equations}

We consider a two--level system with ground state $\ket{g}$ and
excited state $\ket{e}$ which is driven by a classical light field and
coupled to a heatbath of radiation modes.
The starting point of the derivation of Mollow's pure state analysis is the
definition of a reduced atomic density operator in the subspace containing
exactly $n$ scattered photons
according to
\be
\rho^{(n)} \targ{t} = \Tr_B \{ \OP{P}^{(n)} \OP{\rho}\targ{t} \} \quad
\mbox{($n=0,1,2,\ldots$)}
\ee
with $\OP{\rho}\targ{t}$ the density operator of the combined atom + field
system, $\OP{P}^{(n)}$ the projection operator onto the $n$-photon subspace,
and $\Tr_B $ indicating the trace over the radiation modes. The probability of
finding exactly $n$ photons in the field is
\be
P^{(n)} (t)  = \Tr_S \rho^{(n)} \targ{t} \equiv \rho^{(n)}_{gg} \targ{t} +
\rho^{(n)}_{ee} \targ{t}
\ee
with
$\Tr_S$ the trace over the atomic variables (the ``system''). It has been
pointed out in Refs. \Cite{M75,BZ,ZMW87} that $\rho^{(n)} \targ{t}$ obeys the
equation of motion
\be \Label{MollowDMnn}
\frac{d}{dt} \rho^{(n)} \targ{t} =
-i \left(
\Heff \rho^{(n)} \targ{t} - \rho^{(n)} \targ{t} \Heff^\dagger
\right)
+ \Gamma \sigma_{-}  \rho^{(n-1)} \targ{t}  \sigma_{+} \; .
\ee
Here
\be  \Label{MollowDMn}
\Heff =
(-\Delta -i \half \Gamma) \sigma_{ee}
- \left (\half \Omega \sigma_+ + H.c. \right)
\ee
is the non--hermitian Wigner-Weisskopf Hamiltonian of the radiatively
damped and driven two--level atom with $\sigma_{+}=\ket{e}\bra{g}$,
$\sigma_{-}=\ket{g}\bra{e}$, $\sigma_z=\ket{e}\bra{e}-\ket{g}\bra{g}=
\sigma_{ee}-\sigma_{gg}$ the Pauli spin
matrices describing a two--level atom, $\Delta$ the laser detuning,
$\Gamma$ the spontaneous decay rate of the upper state of the
two-level system and $\Omega$ the Rabi frequency. Summing over the $n$--photon
contributions
\be
\rho \targ{t} = \sum_{n=0}^{\infty}  \rho^{(n)} \targ{t}
\ee
with $\rho (t)$ the reduced atomic density operator,
Eq.~(\Ref{MollowDMnn}) reduces to the familiar optical
Bloch equations (OBEs)
\be \Label{MollowOBE}
\frac{d}{dt} \rho \targ{t} =
-i \left(
\Heff \rho \targ{t} - \rho \targ{t} \Heff^\dagger
\right)
+ \Gamma \sigma_{-} \rho \targ{t}  \sigma_{+} \; .
\ee
The $n$--photon density matrix $\rho^{(n)} \targ{t}$ is seen to obey a
hierarchy of equations where the $n-1$ photon density matrix provides
the feeding term for the $n$--photon term, $ \ldots \rightarrow
\rho^{(n-1)}\rightarrow \rho^{(n)} \ldots$. By formal integration of this
hierarchy we obtain
{\em Mollow's pure state representation} of the atomic density matrix
\bea \Label{Mollowrho}
\rho \targ{t} 
=
\ket{\pct  \targ{t}} \bra{\pct \targ{t}}
&&+  \sum_{n=1}^{\infty}
\int_{0}^{t} dt_n \int _{0}^{t_n} dt_{n-1} \ldots \int _{0}^{t_2} dt_1 \\ &&
\ket{\pct  \targ{t|t_n,t_{n-1},\ldots,t_1}} \bra{\pct \targ{t|t_n,t_{n-1},
\ldots,t_1}}
\eea
with atomic wave functions obeying the equation of motion
\be \Label{Mollowprop}
\frac{d}{dt} \ket{\pct  \targ{t|t_n,\ldots,t_1}}
= - i \Heff \ket{\pct  \targ{t|t_n,\ldots,t_1}} \quad 
\mbox{($t \ge t_n \ge t_{n-1} \ge\dots  \ge t_1$)}
\ee
and defined recursively by the jump condition
\be \Label{Mollowqj}
\ket{\pct  \targ{t_n|t_n,t_{n-1},\ldots,t_1}}
=
\sqrt{\Gamma} \sigma_- \ket{\pct  \targ{t_n|t_{n-1},\ldots,t_1}}
\ee
with $\norm{\pct (t=0)}=1$.
Eq.~(\Ref{Mollowrho}) gives the atomic density matrix in terms of {\em
pure atomic states} $\pct \targ{t|t_n,\ldots,t_1}$. 
\subsubsection{Interpretation}
The essential features 
behind this construction are as follows. According to Mollow \Cite{M75} Eq.~(
\Ref{Mollowrho}) can be
interpreted as the time evolution of an atom in a time interval
$(0,t]$ which emits exactly $n$ photons at times $t_1,t_2,\ldots,t_n$.
According to (\Ref{Mollowqj}) atoms in the ground state with $n$
photons in the scattered field are created at a rate $\Gamma \norm{
\sigma_- \pct \targ{t|t_{n-1},\ldots,t_1}}^2$.  Each photon emission is
accompanied by a reduction of the atomic wave function to the ground
state $\ket{g}$ as described by the operator $\sigma_-$ in
Eq.~(\Ref{Mollowqj}). This is what we call a {\em quantum jump}. The
time evolution between the photon emissions is governed by the
non--hermitian Wigner--Weisskopf Hamiltonian $\Heff$ which describes
the reexcitation of the radiatively damped atom by the laser field.

As has been pointed out in Ref.~\Cite{ZMW87} in the context of
discussion of quantum jumps in three-level systems \Cite{BZ,qjexp},
Eq.~(\Ref{Mollowrho}) has the structure expected from the Srinivas
Davies theory of continuous measurement for counting processes
\Cite{SD81}, and in particular supports the interpretation of \be
p_0^t(t_n,\ldots,t_1) = \norm{ \pct \targ{t|t_n,\ldots,t_1} }^2 \ee as
an {\em exclusive probability density} that the atom emits $n$ photons
a times $t_1,t_2,\ldots,t_n$ (and no other photons) in the time
interval $(0,t]$. Thus the time evolution can be simulated:
Fig.~\Ref{fig:Mollow1} shows the excited state population
$\abs{\ME{e}{\psi_c(t)}}^2$ with normalized wavefunction $\psi_c
\targ{t} = \pct \targ{t} / \norm{ \pct \targ{t}}$, corresponding to a
single run as a function of time for a Rabi frequency $\Omega =
\Gamma$ and detuning $\Delta=-\Gamma$. The decay times (indicated by
arrows) and quantum jumps where the atomic electron returns to the
ground state are clearly visible as interruptions of the Rabi
oscillations. After a quantum jump the atom reset is to the ground state,
and it then is reexcited by the laser: in the photon statistics of the
emitted light this leads to antibunching, i.e. two photons will not be
emitted at the same time \Cite{WallsMilburn}. Averaging over these
trajectories gives the familiar transient solution of the optical
Bloch equation. Although Mollow's derivation and somewhat intuitive
interpretation was restricted to a specific example, the basic
structure and ideas which emerge from his analysis are valid in a much
more general context which is what we will be concerned with in the
following sections.
\begin{figure}
\begin{center}\
\psfig{file=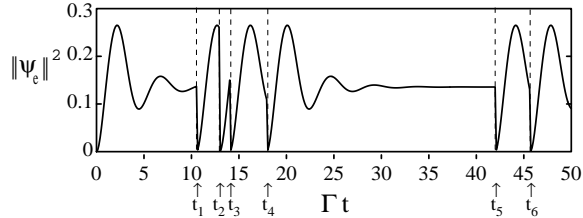,width=3in}
\end{center}
\caption{Plot of a realization of the Monte Carlo wavefunction as a
function of time: excited state probability $\abs{\ME{e}{\psi_c(t)}}^2$
{\protect\Cite{DRZ92}}.
}
\Label{fig:Mollow1}
\end{figure}

\section{Wave-function quantum stochastic differential equations}

\subsection{The Model}

The standard model of quantum optics \Cite{QN,B90} considers a  system
interacting with a heatbath consisting of many harmonic oscillators
representing the
electromagnetic field. The Hamiltonian for the combined system is
($\hbar=1$)
\be
H_{\rm tot} = \Hsys + \HB + \Hint
\ee
with
\be \Label{HB}
\HB = \int \, d\omega \, \omega \,b(\omega)^\dagger b(\omega)
\ee
the Hamiltonian for the heatbath, and $b(\omega)$ the annihilation
operator satisfying the canonical commutation relations (CCR)
\be  \Label{CCR}
[b(\omega),b(\omega)^\dagger]=\delta(\omega - \omega') \; .
\ee
For simplicity, we assume only a single heatbath. Generalization to
many reservoirs represents no conceptual difficulty.  The interaction
Hamiltonian (\Ref{Hint}) is based on a linear system -- field coupling
in a rotating wave approximation (RWA),
\be \Label{Hint}
\Hint = i \frac{1}{\sqrt{2\pi}} \int \, d\omega \, \kappa(\omega) [c b(
\omega)^\dagger- b(\omega) c^\dagger]
\ee
with $c$ a system operator (the ``system dipole''), and
$\kappa(\omega)$ coupling functions. The system Hamiltonian $\Hsys$ is
left unspecified.

\subsubsection{Approximations}
Quantum noise theory is based on the following approximations \Cite{QN}: (i)
 Rotating wave approximation (RWA) and smooth system--bath coupling, and
(ii) a Markov (white noise) approximation.

\begin{itemize}
\item {\em  Rotating wave approximation:}
For simplicity, we assume that for the bare (interaction free) system the
system
dipole $c$ evolves as $c(t) = c e^{-i\omega_0 t}$ with $\omega_0$ the
resonance frequency of the system. Thus the system will be dominantly
coupled to a band of frequencies centered around $\omega_0$. Validity of the
RWA
requires that the frequency integration in $\Hint$ is restricted to a
range of frequencies
\be \Label{freqinterval}
\int_{\omega_0 - \vartheta}^{\omega_0+\vartheta} \, d\omega \ldots \,
\ee
with cutoff $\vartheta \ll \omega_0$. This assumes a separation of
time scales: the optical frequency $\omega_0$ is much larger than
the cutoff $\vartheta$ which again is much larger than the typical
frequencies of the system dynamics and frequency scale induced by the
system -- bath couplings (decay rates). Furthermore, we assume a
smooth system--bath coupling $\kappa(\omega)$ in the frequency
interval (\Ref{freqinterval}): we set $\kappa(\omega) \rightarrow
1
$ (a constant factor can always be reabsorbed in a definition of $c$).

We transform to the interaction picture with respect to the free
dynamics of the ``bare system + field.'' In the interaction
Hamiltonian (\Ref{Hint}) this amounts to the replacements $c
\rightarrow c e^{-i \omega_0 t}$, $b(\omega) \rightarrow b(\omega)
e^{-i \omega t}$, so that in the interaction picture the interaction
Hamiltonian becomes
\be
\Hint^{(\vartheta)} (t)=  i \gs{}  [ b^{(\vartheta)}(t)^\dagger c - b^{(
\vartheta)}(t) c^\dagger ]
\ee
with
\be
b^{(\vartheta)}(t) = \frac{1}{\sqrt{2 \pi}} \int_{\omega_0 -
\vartheta}^{\omega_0+\vartheta} \, d\omega \; b(\omega) e^{-i (\omega
- \omega_0)t} \, .
\ee
The time evolution operator $U^{(\vartheta)}\targ{t}$ from time $0$ to
$t$ in the interaction picture obeys the Schr\"odinger equation
\be \Label{Uint}
\frac{d }{dt} U^{(\vartheta)}\targ{t} = -i (H + \Hint^{(\vartheta)} (t)) U^{(
\vartheta)}\targ{t}
\ee
with $H$ the transformed system Hamiltonian. Note that by going to the
interaction picture ``fast'' optical frequencies $\omega_0$ have
disappeared (``transformation to a rotating frame'').
\item {\em The  Markov approximation or white noise approximation}:
this consists
of taking the limit $\vartheta
\rightarrow \infty$ in Eq.~(\Ref{Uint}),
\be \Label{BoseField}
b^{(\vartheta)}(t) \rightarrow b(t) : = \frac{1}{\sqrt{2 \pi}}
\int_{-\infty}^{\infty} \, d\omega b(\omega) e^{-i (\omega -
\omega_0)t} \, .
\ee
so that the commutator for the fields $b(t)$ in the time domain
acquires a $\delta$--function form,
\be \Label{comm}
[b(t), b(t')^\dagger] = \delta (t-t') \; .
\ee
The operator $b(t)$ is a driving field for the equation of motion
(\Ref{Uint}) at time $t$. In the following we should interpret the
parameter $t$ to mean {\em the time at which the initial incoming
field will interact with the system}, rather then specifying that
$b(t)$ is a time-dependent operator at time $t$.

\end{itemize}

\subsection{Quantum Stochastic Calculus}
\Label{sec:QSC}
The commutator (\Ref{comm}) is a $\delta$--function because of the
Markov approximation. The Markovian equations that result have a
greatly simplified form, but this simplification arises at the expense
of having to define stochastic calculus, leading to the concepts of Ito and
Stratonovich integration much as in the classical case. We give here a
heuristic review of these ideas.

Quantum stochastic calculus (QSC) is a non-commutative analogue of
Ito's stochastic calculus. It was developed originally as a
mathematical theory of quantum noise in open system and more recently
found application to measurement theory in quantum mechanics.

For simplicity we consider the situation in which the field is in a
vacuum state, so that $b (t) \ket{ \rm vac} =0$, and thus $\AV{b(t)
b^\dagger (t')}=\bra{\mbox{\rm vac}} b(t) b^\dagger(t')\ket{ \mbox{\rm vac}} =
\delta (t-t')$ and $\AV{b^\dagger (t) b(t')}=0$.  We define
\be
B\targ{t} : = \int_0^t \, ds \, b(s), \quad B\targ{t}^\dagger : =
\int_0^t \, ds
\, b(s)^\dagger \; .
\ee
For the vacuum averages these definitions lead to
\be   \Label{eq:2.21}
\AV{[B\targ{t} - B\targ{t_0}] [ B\targ{t}^\dagger -  B\targ{t_0}^\dagger]} =
\abs{ t - t_0} %
\\
\ee
\be
\AV{B\targ{t} - B\targ{t_0}}  = \AV{ [B\targ{t}^\dagger - B\targ{t_0}^
\dagger]^2 } = \AV{ [B\targ{t} - B\targ{t_0}]^2 } =0 \, .
\ee
A
quantum stochastic calculus of the Ito type, based on the increments
\be \dB{t} = B\targ{t\!+\!dt} - B\targ{t}, \quad \dB{t}^\dagger =
B\targ{t\!+\!dt}^\dagger - B\targ{t}^\dagger \ee and $dt$ has been
developed by Hudson and Parthasarathy
\Cite{HP}. The pair $B\targ{t}$, $B\targ{t}^\dagger$ are the non--commutative
analogues of complex classical  Wiener processes.

\subsubsection{Quantum stochastic integration}
We consider two definitions of quantum stochastic integration: {\em
Ito},
\be \Label{ItoInt}
\Ii \int_0^t \, f(s) \, \dB{s}
=
\lim_{n\rightarrow \infty}
\sum_{i=0}^n
f(t_i) [B\targ{t_{i+1}} - B\targ{t_i}]
\ee
and {\em Stratonovich}
\be \Label{StraInt} \Label{eq:2.23}
\Si \int_0^t \, f(s) \, \dB{s}
=
\lim_{n\rightarrow \infty}
\sum_{i=0}^n
\half[ f(t_{i+1}) +f(t_{i}) ] [B\targ{t_{i+1}} - B\targ{t_i}] \, .
\ee
In both cases $f(t)$ is a {\em nonanticipating} (or {\em adapted})
function, i.e. an operator valued quantity which depends
only on $B\targ{s}$ {\em etc.} for $s<t$. There are analogous
definitions for the Ito and Stratonovich versions of
\be
\int_0^t \, f(s) \, \dB{s}^\dagger,
\int_0^t \, \dB{s} \,  f(s) ,
\int_0^t \, \dB{s}^\dagger \,  f(s).
\ee
The basic difference between the Ito and Stratonovich form of the
integrals is that in the Ito form the term $f(t_i)$ and
$[B\targ{t_{i+1}}-B\targ{t_i}]$ are independent of each other, whereas
in the Stratonovich form $[ f(t_{i+1}) +f(t_{i}) ]$ is not independent
of $[B\targ{t_{i+1}}-B\targ{t_i}]$.

\subsubsection{Example and discussion}
If we use the property (\Ref{eq:2.21}),
we
find
\begin{equation}\Label{eq:2.25}
\AV{ \Ii \int_{0}^{t}\, \dB{s}B\targ{s}^\dagger }
= 0
\end{equation}
\begin{equation}\Label{eq:2.26}
\AV{ \Si \int _{0}^{t} \, \dB{s} B\targ{s}^\dagger
 } =
\frac{1}{2}|t|\,.
\end{equation}
In fact this example shows the three main principles involved.
\begin{enumerate}
\item[i]{\em Stratonovich integration follows the
rules of conventional calculus.} For example, the conventional
differential of $B(t)B\targ{t}^\dagger$ is the Stratonovich
differential
\begin{equation}\Label{eq:2.27}
({\rm S})\,d(B\targ{t} B \targ{t}^\dagger ) = \, \dB{t}
B\targ{t}^\dagger + B\targ{t}
\, dB \targ{t}^\dagger
\end{equation}
so that
\begin{equation}\Label{eq:2.28}
\left . B\targ{t} B\targ{t}^\dagger \right | _{0}^{t}
 = \Si \int \limits _{0}^{t} \, \dB{s} B\targ{s}^\dagger + \Si
\int_{0}^{t} B\targ{s} \, \dB{s}^\dagger
\end{equation}
and using (\Ref{eq:2.21},\Ref{eq:2.23}) and taking averages, we find
\begin{equation}\Label{eq:2.29}
|t| = {1\over2}|t| + {1\over2}|t|
\end{equation}
as should be the case.
\item[ii]{\em  Ito Integration increments
are independent of and commute with the integrand.} From this
independence we see that we can factorize the average so that
\begin{equation}\Label{eq:2.30}
\AV{ \Ii \int_0^t \, \, \dB{s} \, B\targ{s}^\dagger } =
\int \AV{\dB{s}} \AV{B\targ{s}^\dagger} = \int_0^t \, 0 = 0
\end{equation}
For Ito differentials the conventional rule (\Ref{eq:2.27}) is
replaced by the rules:
\begin{enumerate}
\item Expand all differentials to {\em second\/}
order.
\item    Use the multiplication rules [for a vacuum
state---otherwise use the rules (\Ref{nonvacuum})]
\begin{eqnarray}\Label{eq:2.31}
 dt^{2} =dB^ {2}\targ{t} =dB\targ{t}^{{\dagger}^{2}} &=& \,
dB^\dagger\targ{t} \, \dB{t} =0
\nonumber %
\\
 \, \dB{t} dt=\, dB\targ{t}^\dagger  \, dt &=&dt \, \, \dB{t} = dt \,
\, dB\targ{t}^\dagger = 0
\nonumber
\\
\dB{t} \, dB\targ{t}^\dagger &=&dt \, .
\end{eqnarray}
\end{enumerate}

Using these rules, we find
\begin{eqnarray}\Label{eq:2.32}
\Ii d \left[ B\targ{t} B^\dagger \targ{t} \right]
&=& \dB{t} B\targ{t}^\dagger + B\targ{t} \, \dB{t}^\dagger + \, \dB{t}
\, \dB{t}^\dagger
\nonumber
\\
&=&\dB{t}\, B\targ{t}^\dagger+B\targ{t} \, \dB{t}^\dagger+dt \, ,
\end{eqnarray}
so that using the Ito integral, we get
\begin{equation}\Label{eq:2.33}
\AV{ B\targ{t} B\targ{t}^\dagger } | _{0}^{t}=|t| =0+0+|t|\,
\end{equation}
in agreement with (\Ref{eq:2.29}).  These rules are sufficient for
ordinary manipulation, but are defined only when the state is the
vacuum.  Rules for non-vacuum states will be discussed briefly at the
end of this section.
\item[iii]{\em The mean value of an Ito integral is always zero.}
This follows from the independence of  $ \, \dB{t} = B\targ{t\!+\!dt}-B
\targ{t}$ and $f(t)$, which has been assumed to be nonanticipating.
\end{enumerate}

As in the case of classical stochastic differential equations, an
equation
\be
\frac{d}{dt} X\targ{t} = F\targ{X\targ{t},t} +   G\targ{X\targ{t},t} \, b(t)  +
H
\targ{X\targ{t},t} \, b^\dagger(t)
\ee
cannot be rigorously considered as a {\em differential equation},
since the terms involving $b(t)$ and $b^\dagger(t)$ are in some sense
infinite. However this equation can be regarded as an {\em integral
equation}
\begin{eqnarray}\Label{intequation}
&&X(t)-X(t_0) =  \int_{t_0}^{t}dX(t') \\
&&= \int_{t_0}^{t}\left\{F\targ{X\targ{t'},t'}\,dt' +
G\targ{X\targ{t'},t'} \,dB(t')  +   H\targ{X\targ{t'},t'} \,dB^\dagger(t')
\right
\}  \nonumber
\end{eqnarray}
As in the case of non-quantum  stochastic differential equations, a simplified
notation is used, in which the explicit symbols of integration are dropped:
\begin{eqnarray}\Label{SDE}
dX(t)
&=& F\targ{X\targ{t},t}\,dt +
G\targ{X\targ{t},t} \,dB(t)  +   H\targ{X\targ{t},t} \,dB(t)^\dagger
\end{eqnarray}
which is known as a {\em stochastic differential equation}.

\subsubsection{Comparison of Ito and Stratonovich Stochastic Differential Equations (SDE)}
The Stratonovich definition is natural for physical situations where
the white noise approximation is an idealization.  While the
Stratonovich form has its merits of satisfying ordinary calculus, the
equation as it stands is rather like an implicit algorithm for the
solution of a differential equation which makes manipulations rather
difficult.

In the Ito definition the differentials $\dB{t} =
B\targ{t\!+\!dt}-B\targ{t}$ {\it etc.} point to the future.  Quantum
stochastic Ito calculus obeys simple formal rules which can be
summarized as follows:
\begin{enumerate}
\item [i] By the CCR (\Ref{CCR}) nonanticipating functions $X\targ{t}$ 
commute with the fundamental differentials, $[X\targ{t},\dB{t}]=0$ {\em etc.}. 
\item[ii] For two functions $X\targ{t}$ and $Y\targ{t}$ satisfying equations  
of the type (\Ref{SDE}), the differential of the product $X\targ{t} Y\targ{t}$ 
is given by
\be \Label{ItoMulti}
d [X\targ{t} Y\targ{t}] = [d X\targ{t}] \, Y\targ{t} + X\targ{t} \,
[dY\targ{t}] + dX\targ{t} \, dY\targ{t}
\ee
where the Ito correction (the last term on the RHS of
Eq.(\Ref{ItoMulti})) has to be computed by means of the Ito table
\be \Label{ItoRules}
 \dB{t} \, \dB{t}^\dagger = dt \quad \mbox{(vacuum state)}
\ee
and all other products involving $d B\targ{t}$, $\dB{t}^\dagger$ and
$dt$ vanish.
\end{enumerate}

\subsection{Quantum Stochastic Schr\"odinger, Density Matrix and Heisenberg 
Equations}

The definition of Ito and Stratonovich integration allows us to give
meaning to a {\em quantum stochastic Schr\"odinger equation} for the
time evolution operator $\OP{U}\targ{t}$ of the system interacting
with Bose fields $b(t)$, $b^\dagger(t)$.  The Schr\"odinger equation
(\Ref{Uint}) must be interpreted as a Stratonovich SDE \be
\Label{SSEStratonovichDE} \Si \, d \OP{U}\targ{t} = \{ -i H \, dt +
\gs{} \dB{t}^\dagger c - \gs{} \dB{t} c^\dagger \} \OP{U}\targ{t} \, .
\ee It is advantageous to convert this Stratonovich equation to Ito
form.  By conversion to Ito form, we can get a modified equation in
which the increments $\dB{t}$, $\dB{t}^\dagger$ and independent of
$ \OP{U}\targ{t} $.  This Ito form of the quantum stochastic
Schr\"odinger equation is
\be \Label{SSEIto}
\Ii \, d \OP{U}\targ{t} =
\{ (-i H - \half \g{} c^\dagger c) \, dt + \gs{} \dB{t}^\dagger \, c -
\gs{} \dB{t}\, c^\dagger \} \OP{U}\targ{t} \quad (\OP{U}\targ{0} =
\one).
\ee
As the solution $\OP{U}\targ{t}$ of (\Ref{SSEIto}) is
nonanticipating, since the increments $\dB{t}$, $\dB{t}^\dagger$ point
to the future, we have $[\OP{U}\targ{t},\dB{t}]=[\OP{U}\targ{t},dB
\targ{t}^\dagger]=0$.

\vspace{2mm}
{\small {\em Remark: Conversion from Stratonovich to Ito form:} this
conversion proceeds analogous to the classical case. Consider the Ito
equation
\be \Label{SSEItoAppendix}
\Ii \, d {\OP{U}\targ{t}} = 
\left\{
\alpha \, dt + \beta \dB{t}^\dagger - \beta^\dagger \dB{t}^\dagger
\right\}
{\OP{U}\targ{t}}
\ee
We consider an arbitrary Stratonovich integral of a function
$\ket{\phi \targ{t}} = \OP{U}\targ{t} \ket{\phi \targ{0}}$ which obeys Eq.~(\Ref{SSEItoAppendix})
\bea
\Si \int \, \dB{t} {\phi \targ{t}} &=& \lim \sum \Delta B_i \half [{\phi
\targ{t_{i+1}}} + {\phi\targ{t_{i}}}] %
\\ 
  &=& \lim \sum \Delta B_i [ {\phi\targ{t_i}} + \half (\alpha \Delta
t_i +
\beta \Delta B_i^\dagger - \beta^\dagger \Delta B_i ){\phi\targ{t_i}} ] \nonumber
\eea
and by $\Delta B_i \, \Delta B_i^\dagger \rightarrow \Delta t_i$
\bea
\Si \int \, \dB{t} {\phi (t)}  &=& \lim \sum
\left\{
\Delta B_i {\phi\targ{t_i}} + \half \beta \Delta t_i {\phi\targ{t_i}}
\right\} %
\\ 
&=&\Ii \int \, \dB{t} \, {\phi \targ{t}} + \half \int \, dt \, \beta
\, {\phi\targ{t}} \nonumber
\eea
and similar expressions for the other integrals. Remembering that
(\Ref{SSEItoAppendix}) is a shorthand for an integral equation
we see that (\Ref{SSEItoAppendix}) is
equivalent to (\Ref{SSEStratonovichDE}) if $\beta = c$, $\beta^\dagger
= c^\dagger$, $\alpha = -i H -\half c^\dagger c$.  }
\vspace{2mm}

The formal solution of the Eq.~(\Ref{SSEIto}) can be written as
\be  \Label{Utt}
\OP{U}\targ{t} = T \, {\rm exp} {\int_0^t \, \left( -i H \, ds + \gs{} \dB{s}^
\dagger c -
\gs{} \dB{s} \, c^\dagger \right)}
\ee
where $T$ denotes the time--ordered product. This equation gives
\bea \Label{Ut}
d \OP{U} \targ{t} & \equiv  & \OP{U}\targ{t\!+\!dt} - \OP{U}\targ{t} =
\left[
{\rm exp} { \left( -i H \, dt + \gs{} \dB{t}^\dagger c - \gs{} \dB{t} \, c^
\dagger
\right) } -1
\right] \OP{U}\targ{t}  \nonumber
\\
&=& \sum_{n=1}^\infty \frac{1}{n!}
\left(
 -i H \, dt + \gs{} \dB{t}^\dagger c - \gs{} \dB{t} \, c^\dagger
\right)^n \OP{U}\targ{t} \; .
\eea
Applying the Ito rules (\Ref{ItoRules}), terms with $n>2$ vanish, the
term with $n=1$ gives $-i H \, dt + \gs{} \dB{t}^\dagger c - \gs{} \dB{t} \,
c^\dagger$, $n=2$ gives $-\half \g{} c^\dagger c \, dt$, and we recover Eq.~(
\Ref{SSEIto}).  Obviously, from Eq.~(\Ref{Utt}) $\OP{U}\targ{t}$ is
unitary.

The state vector $\ket{\Psi \targ{t} } = \OP{U}\targ{t}
\ket{\Psi\targ{0}} $ with initial condition
$\ket{\Psi\targ{0}}=\ket{\psi} \otimes \ket{\rm vac}$ obeys a
stochastic Schr\"odinger equation
\be \Label{SSE}
 \Ii \, d \ket{\Psi\targ{t}} =
\{
 (-i H - \half \g{} c^\dagger c) \, dt + \gs{} \dB{t}^\dagger \, c - \gs{} 
\dB{t}\,
c^\dagger
\}
\ket{\Psi\targ{t}} \, .
\ee
Since $\ket{\Psi\targ{0}}$ contains the vacuum of the
electromagnetic field, $b(\omega) \ket{\Psi\targ{0}} =0$ and thus $
\dB{t} \ket{\Psi\targ{0}} =0$.  Because $\OP{U}\targ{t}$ commutes with
$\dB{t}$ it follows that $\dB{t} \ket{\Psi\targ{t}} = 0$. This means,
as far as $\dB{t}$ is concerned, $\ket{\Psi\targ{t}}$ is in the vacuum
state and therefore the Ito equation (\Ref{SSE}) can be simplified to
\be \Label{QSSEvac}
 \Ii \, d \ket{\Psi\targ{t}} =
\{
 - i \Heff \, dt + \gs{} \dB{t}^\dagger\, c
\}
\ket{\Psi\targ{t}}
\ee
with
\be
 \Heff = H - i \half \g{} c^\dagger c \, .
\ee
This equation has the following physical interpretation. The term
$\dB{t}^\dagger$ involves the incoming radiation field evaluated in
the immediate future of $t$. Thus this field is not affected by the
system. However, the system does create a self--field which causes the
radiation damping by reacting back on the system. This is the meaning
of the term $ - \half \g{} c^\dagger c \ket{\Psi\targ{t}}$. The
Stratonovich equation does not have this term because the evaluation
of $\dB{t}^\dagger$, half in the future, half in the past, itself
generates the radiation reaction.


\subsubsection{Equation of motion for the stochastic density operator}
From the Schr\"odinger equation (\Ref{SSEIto}), we can derive the
equation of motion for a {\em stochastic density operator} ${\hat
\rho}\targ{t} = \ket{\Psi\targ{t}} \bra{\Psi\targ{t}}$ for the combined system 
+ heatbath as
\bea \Label{SDM}
d \OP{\rho}\targ{t}&=& \OP{U}\targ{t\!+\!dt,t} \OP{\rho}\targ{t}
\OP{U}\targ{t\!+\!dt,t}^\dagger -{\hat
\rho}\targ{t} %
\\  \nn &= & -i\left( \Heff \OP{\rho}\targ{t} - \OP{\rho}\targ{t}
\Heff^\dagger \right) \, dt
 + \g{}  \dB{t}^\dagger \, c \,\OP{\rho}\targ{t}\, c^\dagger \dB{t}%
\\  && +
\gs{} \dB{t}^\dagger c \, {\hat
\rho}\targ{t} + \OP{\rho}\targ{t} \, \gs{} \dB{t} \, c^\dagger \; . \nn
\eea
Here
\be \Label{U}
\OP{U}\targ{t,t_0} \equiv \OP{U}\targ{t} \OP{U}\targ{t_0}^\dagger
\ee
 is the time evolution operator from $t_0$ to $t$. Eq.~(\Ref{SDM}) is
derived by expanding $\OP{U}\targ{t,t\!+\!dt}$ to second order in the
increments (compare Eq.~(\Ref{SSEIto})), and using the Ito rules
(\Ref{ItoRules}).  If we now trace over the bath variables, this will
perform an average over the $\dB{t} $, $\dB{t}^\dagger$ operators. We
use the cyclic properties of the trace, and derive the usual master
equation for $\rho\targ{t} = \Tr_B
\{ \OP{\rho}\targ{t} \}$,
\bea \Label{DM}
\frac{d}{dt} \rho\targ{t} &=&
-i \left( \Heff \, \rho - \rho \, \Heff^\dagger \right) + \g{}
\Jc \rho 
\\ 
&=&\Lc \rho \nn
\eea
with $\Lc$ a Liouville  and ``recycling operator''
\be \Label{recyclingop}
\Jc \rho = c \, \rho \, c^\dagger \; .
\ee

For a system operator $a$ we define a Heisenberg operator $a\targ{t}=
\OP{U}\targ{t} ^\dagger \, a \, \OP{U}\targ{t}$ that obeys the Ito
{\em quantum Langevin equation}
\bea \Label{QuantumLangevinEquation}
da \targ{t} & \equiv & \OP{U}\targ{t\!+\!dt,t}^\dagger \, a\targ{t} \,
\OP{U}\targ{t\!+\!dt,t} - a\targ{t} %
\\  &=& -i [a, H \, dt +i \gs{} \dB{t}
c^\dagger -i \gs{} \dB{t}^\dagger c] + \g{} ( c^\dagger a c - a \, \half
c^\dagger c - \half c^\dagger c \, a ) \, dt \; . \nn
\eea

\subsubsection{Remarks}
\begin{Discussion}
\item
{\em Generalization to more channels:} For reference in later
sections we need the master equation for $\Nd$ output channels. It is
given by
\bea \Label{CPmastereq} \Label{masterequation}
{\dot \rho}(t) &= & -i[H,\rho] +
\sum_{j=1}^\Nd 
(\cop{j} \rho \cop{j}^\dagger
- \half \cop{j}^\dagger \cop{j} \rho -\rho \half \cop{j}^\dagger
\cop{j} )%
\\  &\equiv& -(\Heff \rho -\rho \Heff^\dagger) +
\sum_{j=1}^\Nd {\cal J}_j
\rho \equiv {\cal L} \rho
\eea
with `effective' (non--hermitian) Hamiltonian
\be
\Heff = H - i \half \sum_{j=1}^\Nd 
 \cop{j}^\dagger \cop{j} \; ,
\ee
and `recycling operators'
\be
{\cal J}_j \rho = 
 \cop{j} \rho \cop{j}^\dagger \quad (j=1,\ldots,\Nd)
\; .
\ee
 A master equation of this
form is derived by coupling a system to $\Nd$ reservoirs with system
operators $\cop{j}$ ($j=1,\ldots,\Nd$).

\item {\em Non--vacuum initial states:}
So far we have assumed a vacuum state (a pure state) in the input
field. The above formalism can be generalized to nonpure initial
states, provided we stay in a white noise situation
\Cite{B87,GardinerCollett,G94}. For a Gaussian field with zero mean values
$\AV{b(t)} = \AV{b(t)^\dagger}=0$,
 and  correlation functions
$\AV{b(t)^\dagger b(t')} = N \delta (t-t')$, $\AV{b(t) b(t')^\dagger}
= (N+1) \delta (t-t')$, and $\AV{b(t) b(t')} = M \delta (t-t')$ where
$N$ is the mean photon number and $M$ is a squeezing parameter ($|M|^2
\le N(N+1)$), it can be shown \Cite{QN,B90,GPZ92}  that the Ito table has to be 
extended according to
\bea
&&dB (t)^\dagger dB(t) = N dt \quad dB (t) dB(t)^\dagger = (N+1) dt \nn  %
\\ 
&&dB (t)\,dB(t) = M dt \quad dB (t)^\dagger \,dB(t)^\dagger = M^{*} dt
 \Label{nonvacuum}
\\  && dB (t) \dt = dB^\dagger(t) = 0 \; . \nn
\eea

\end{Discussion}

\subsection{Number Processes and Photon Counting}
The description of photon counting requires an extension of quantum
stochastic calculus to bring in the so--called gauge or number
processes. The number process arises by noting that the total number
of photons counted between time $0$ and $t$
is
\be \Label{defnumberprocess}
\Lambda \targ{t} := \int_0^t \, ds \, b^\dagger(s) b(s) \; .
\ee
This assumes a perfect photodetector with unit efficiency and instantaneous $
\delta$--function response.
The operator $\Lambda\targ{t}$ leads to the construction of the
quantum analogues of a Poisson process. Physically, $\dL{t}$ is an
operator whose eigenvalues are the number of photons counted in the time
interval $(t,t+dt]$.  It is only possible to make sense of $\dL{t}$ in
situations in which the initial state is a vacuum. The reason for this
is physically quite understandable as a white noise field would lead to an
infinite number of counts in any finite time interval.

We will also need Ito rules for $\dL{t}$ \Cite{B90},
\bea \Label{ItoRules2}
d \Lambda\targ{t} \, \dL{t} &=& d \Lambda\targ{t}, \quad \dB{t} \,d
\Lambda\targ{t} =
\dB{t}\, \\ d \Lambda\targ{t} \, 
\dB{t}^\dagger &=& \dB{t}^\dagger
\eea
together with (\Ref{ItoRules}) and all the other products involving
$\dB{t}$, $\dB{t}^\dagger$, $\dL{t}$ and $dt$ vanish.

\vspace{2mm}
{\small We can deduce these Ito rules by noting that for any optical
field with no white noise component, the probability of finding
more than one photon in the time interval $ dt$ will go to zero at
least as fast as $ dt^2$.  We can use this fact to show that if we
discretize the integral $
\ket{A}= \int[d\Lambda (t)^2 - d\Lambda (t)]\ket{\phi\targ{t}},
$
then in the limit of infinitely fine
discretization the mean of the norm of $ \ket{A}$ goes to zero. This
means that in a mean-norm topology, $ \ket{A}\to 0$, and thus we can formally 
write inside
integrals $d\Lambda (t)^2 = d\Lambda (t)$.  This equation ensures
that the only eigenvalues of $ d\Lambda(t)$ are $ 0$ and $ 1$, and
that we can only count either one or no photons in a time interval $
dt$.  }
\vspace{2mm}

\subsection{Input and Output}
\Label{sec:inout}

In quantum optics the Bose fields (\Ref{BoseField}) plays the role of
the electromagnetic field. These fields represents the input and
output channels through which the system interacts with the outside
world.  In particular, $b(t) \equiv b^\inp (t)$ is the {\em input}
field which {\em is the field before the interaction with the system}.
In the work of Gardiner and Collett \Cite{QN,GardinerCollett}, and
Barchielli \Cite{B86,B87,B90} the Heisenberg equations for system
operators are expressed as operator Langevin equations with driving
terms $b(t)$. Gardiner and Collett also define {\em output} operators
for the fields after the interaction with the system,
\be
b^\out (t) := \frac{1}{\sqrt{2 \pi}}
\int_{-\infty}^{\infty} \, d\omega b(\omega,T) e^{-i (\omega -
\omega_0) (t-T)}  
\ee
with $ b(\omega,T)$ the Heisenberg operator of $b(\omega)$ evaluated
at some $T$ in the distant future.  In Ref.~\Cite{GardinerCollett} it
is further shown that
\be \Label{out=in+c}
b^\out (t) = b^\inp (t) + \gs{} c\targ{t}
\ee
which expresses the out field as the sum of the in field and the field
radiated by the source.  The fundamental problem of quantum optics
consists in calculating the statistics of the out field for a system
of interest, in particular the normally order field correlation
functions
\be
\AV{
b^\out (t_1)^\dagger \ldots b^\out (t_n)^\dagger b^\out (t_{n+1})
\ldots b^\out (t_{n+m}) } \;,
\ee
given (all) correlation functions for the in field.

Formally, we introduce {\em output processes} \Cite{B90}
\be \Label{outoperators}
B^\out\targ{t} : = \OP{U}\targ{t}^\dagger \, B\targ{t} \,
\OP{U}\targ{t}, \quad \Lambda^\out\targ{t} : =
\OP{U}\targ{t}^\dagger \, \Lambda\targ{t} \, \OP{U}\targ{t},
\ee
and the output fields by formal (forward) derivatives
\be
b^\out (t) := d B^\out\targ{t}/dt \equiv \lim_{h \rightarrow 0^+}
[B^\out \targ{t+h} - B^\out\targ{t}]/h \; .
\ee
Similar definitions hold for $B^\out \targ{t}^\dagger$ and $b^\out
(t)^\dagger$.  From Eqs.~(\Ref{Uint},\Ref{U}) we see that the time
evolution operator $\OP{U}\targ{t,s}$ from time $s$ to time $t$ only
depends depends only on the fields $b(r)$, $b(r)^\dagger$ with $r$
between $s$ and $t$. By the commutation relations
\be \Label{outoperatorproperties1}
[\OP{U}\targ{t,s},B\targ{\tau}]=0 \quad \mbox{for $\tau \le s$}
\ee
 this gives
\bea \Label{outoperatorproperties}
B^\out \targ{t} &= &\OP{U}\targ{t}^\dagger \OP{U}\targ{T,t} ^\dagger B
\targ{t} \OP{U}\targ{T,t} \OP{U}\targ{t}
\nonumber
\\
& =& \OP{U}\targ{T}^\dagger B
\targ{t} \OP{U}\targ{T} \quad (T\ge t).
\eea

A QSDE for the output processes is derived using QSC \Cite{B90}:
\bea
dB^\out \targ{t} & \equiv & B^\out \targ{t\!+\!dt} - B^\out \targ{t}
\\
& = & \OP{U}\targ{t\!+\!dt} ^\dagger B \targ{t\!+\!dt} \OP{U}\targ{t\!+\!dt} -
\OP{U}\targ{t} ^\dagger B \targ{t} \OP{U}\targ{t}
\nn %
\\
&=&
 \OP{U}\targ{t} ^\dagger  \OP{U}\targ{t\!+\!dt,t}^\dagger
dB \targ{t} \OP{U}\targ{t\!+\!dt,t} \OP{U}\targ{t}
\nn %
\\ \Label{dBout}
&=& dB \targ{t} + \gs{} c\targ{t} \dt
\eea
where the second line follows from (\Ref{outoperators}), the third
line from (\Ref{outoperatorproperties1}), and the last line is
obtained from (\Ref{Uint}) and the Ito rules (\Ref{ItoRules}) in
agreement with Eq.~(\Ref{out=in+c}).  In a similar way one finds
\bea
\\
&& d\Lambda^\out\targ{t} = \dL{t} + \gs{} c\targ{t} \, \dB{t}^\dagger +
\gs{} c\targ{t}^\dagger \, \dB{t} + \g{} c\targ{t}^\dagger c\targ{t} \, dt
\Label{dLambdaout} \; .
\eea
which expresses the operator for the count rate for the out field as the
sum of the in operator, an interference term between the source and in
field, and the direct system term.

\section{Counting and Diffusion Processes: Stochastic Schr\"odinger Equation
and Wave Function Simulation}

\subsection{Photon Counting and Exclusive Probability Densities}

\Label{PhotonCounting}\Label{sec:PhotonCounting}

In this section we discuss photon counting of the output field as
realized by a photoelectron counter \Cite{QN,Carmichael} (see Fig.~\ref{fig:ch}a).
Mathematically, this corresponds to a class of continuous measurements
known as ``counting processes'' \Cite{BB91}.
\begin{figure}
\begin{center}\
\psfig{file=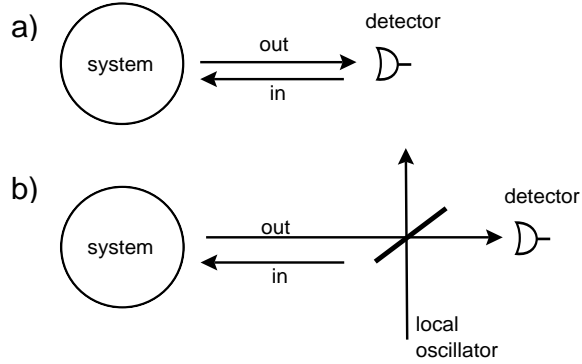,width=3in,angle=90}
\end{center}
\caption{ (a) Photon counting, and (b) homodyne detection.
}\Label{fig:ch}
\end{figure}
We consider a system which is coupled to $j=1,\ldots,\Nd $ output
channels with counters which act continuously to register arrival of
photons. A complete description of counting process is given by the
family of {\it exclusive probability densities} (EPDs):
\begin{ListNoNumber}
\item
$P_{t_0}^t (0| \rho)$ = probability of having no count in the time
interval $(t_0,t]$ when the system is prepared in the state $\rho$ at
time $t_0$.
\item
$p_{t_0}^{t}(j_m,t_m;\ldots;j_1,t_1|\rho)=$ multitime probability
density of having a count in detector $j_1$ at time $t_1$, a count in
detector $j_2$ at time $t_2$ {\it etc.} and no other counts in the
rest of the time interval $(t_0,t]$ (with $j_k=1,\ldots,\Nd $; $t_0 <
t_1< \ldots < t_m \le t$).
\end{ListNoNumber}
Knowledge of EPDs allows the reconstruction of the whole counting
statistics.
\subsubsection{Mandel's counting formula}
Physically, the operator for the output current of a photodetector is
proportional to the photon flux. To simplify notation we restrict
ourselves in the following to the case of a single output channel $\Nd
=1$. Furthermore, since the aim of the present discussion is to
discuss the relation with the stochastic Schr\"odinger equation, we
will concentrate on the case of ideal photodetection with unit
efficiency.  The photon flux operator is
\be
d\Lambda^\out\targ{t} \equiv
b^\out(t)^\dagger b^\out (t) \dt \,=:\,
{\hat I}(t) \dt  \; .
\ee
Here $\Lambda^\out \targ{t}$ is the operator corresponding to the number of
photons up to time $t$ in the out field. We note that the operators
$\Lambda^\out \targ{t}$ are a family of commuting selfadjoint
operators.
Photoelectric detection realizes a measurement of
the compatible observables $\Lambda^\out \targ{t}$.
In quantum optics the usual starting point is the photon counting
formulas first derived by Kelley and Kleiner, Glauber and Mandel
\Cite{Carmichael}:
\be \Label{KK}
P_m(t,t_0) =
\AV{:
\frac{1}{m!} \left( \int_{t_0}^{t}ds \,{\hat I}(s)  \right)^m
\exp \left(-  \int_{t_0}^{t}ds \,   {\hat I}(s)\right)
:} \;
\ee
gives the probability for $m$ photoncounts in the time interval
$(t_0,t)$ where $:\ldots :$ indicates normal ordering, and
$\AV{\ldots} = \Tr_{S \oplus B} \{
\ldots \OP{\rho}_0 \}$ with $\OP{\rho}_0 = \rho \otimes \ket{\rm
vac} \bra{\rm vac}$ is the quantum expectation value with respect to
the state of the system.  Eq.~(\Ref{KK}) is valid for a detector with
unit efficiency.  In particular, the probability of no count in a time
interval $(t_0=0,t]$ is
\be \Label{PF}
P_{t_0=0}^{t}(0| \rho)=
\AV{ : \exp\left(- \int_{0}^{t} ds  \OP{I}(s) \right):  } \; ,
\ee
and the multitime probability density for detecting (exactly) $m$ photons at
times $t_1,\ldots,t_m$ in the time interval $(t_0=0,t]$ is
\be \Label{pF}
p_{t_0=0}^{t}(t_m,\dots,t_1 | \rho)=
\AV{ : \OP{I}(t_1) \ldots  \OP{I}(t_m) \exp\left(- \int_{0}^{t} ds  \OP{I}(s)
\right):  } \; .
\ee
This establishes the relation to the EPDs introduced above.

On the
other hand, the {\em  non--exclusive multitime probability density} for
detecting a first
photon at time $t_1$, a second (not necessarily the next) photon at
time $t_2$ {\em etc.} and the $m$--th photon at $t_m$ is given by the intensity
correlation function
\be \Label{IntensityCorrelation}
\AV{ : \OP{I}(t_1) \ldots  \OP{I}(t_m):  } \; .
\ee

\subsubsection{The characteristic functional and system averages}
\Label{sec:CF}
As a next step we wish to express the out field correlation functions
in terms of system averages.  The standard approach is to express in
Eqs.~(\Ref{PF}) and (\Ref{pF}) the out field in terms of the in field
plus the source contribution, $b^\out (t) = b^\inp (t) + c\targ{t}$,
and to prove and apply the {\em quantum fluctuation regression theorem }
(QFRT) to evaluate the normally- and time-ordered multitime
correlation function for the source operator $c \targ{t}$ \Cite{QN}.
Instead we will follow here the elegant procedure outlined by
Barchielli \Cite{B87,B90} of defining and evaluating a characteristic
functional for the EPDs.

We define a characteristic functional as the expectation value of the 
characteristic operator $\OPVout_t [k]$ according to
\bea \Label{CFV}
\Phi_t[k] &= &
\AV{ \OPVout_t [k] } \quad \mbox{with} \quad \OPVout_t[k] :=
\exp \left( i \int_0^{t}   k(s) \,  \OP{I}(s) \, ds     \right) \; .
\eea
When we use the normal ordering relation \Cite{Louisell}
\bea
 \exp \left( i \int_0^{t} k(s) \, \OP{I}(s) \, ds \right)  = 
: \exp \left( \int_0^{t} (e^{i k(s)}-1)\, \OP{I}(s)
\, ds \right) :\; , 
\eea
expand the exponential, and use expressions (\Ref{PF}) and (\Ref{pF})
we obtain
\bea \Label{defCF}
\Phi_t[k] := P_0^t(0|\rho)  +  \sum_{m=1}^{\infty}  && \int_{0}^{t} dt_m 
\int_{0}^{t_m} dt_{m-1}
\ldots  \int_{0}^{t_2} dt_1 
\nonumber \\ &&\exp \left( i \sum_{r=1}^{m} k(t_r) \right) 
p_{0}^{t}(t_m,\dots,t_1 | \rho)
\eea
which relates the characteristic functional to the exclusive
probability densities. Thus $\Phi_t[k]$ determines uniquely the whole counting
process up to time $t$.  Functional differentiation of $\Phi_t[k]$
with respect to the test functions $k(t)$ gives the moments,
correlation functions and counting distributions.

According to Eq.~(\Ref{outoperators}) the out and in processes are
related by
$\Lambda^\out \targ{t}= U\targ{t}^\dagger \Lambda \targ{t}
U\targ{t}$ and the characteristic operator satisfies the QSDE
\bea \Label{dVout}
&& d\OPVout_{t}[k] = \OPVout_{t}[k] \, \left(  e^{i k(t)} -1 \right) d\Lambda^
\out\targ{t}
 \quad (\OPVout_0[k]=\one),
\\
&& d\Lambda^\out\targ{t} = \dL{t} + \gs{} c\targ{t} \, \dB{t}^\dagger +
\gs{} c\targ{t}^\dagger \, \dB{t} + \g{} c\targ{t}^\dagger c\targ{t} \, dt \; .
\eea
In a similar way we can express the characteristic operator for the out field
in terms of the in field;
\bea
\OPVout_t[k] &=& U\targ{t}^\dagger
\exp \left( i \int_0^{t}   k(s) \, d\Lambda \targ{s} \right)
 U\targ{t} %
\\
             &\equiv& U\targ{t}^\dagger \OPV_t[k] U\targ{t} \; . \nn
\eea
This allows us to write
\bea \Label{chidef}
\Phi_t[k] &=& \Tr_{S \oplus B} \{ \OPV_t[k]\, U\targ{t} \OP{\rho}(0) U\targ{t}^
\dagger \} %
\\
&=& \Tr_S {\chi}_t[k] \quad \mbox{with} \quad {\chi}_t[k] := \Tr_B
\{ \OPV_t[k]\, \OP{\rho}(t) \} \; .
\eea
which expresses the characteristic functional as the system trace over a
characteristic density-like operator $\chi _t [k]$ in the system space.
Using QSC we can derive an equation for $\OPV_t[k]\, \OP{\rho}(t)$,
\be \Label{Itocountoperator}
d [ \OP{V}_t[k] \; \OP{\rho}(t)] = d \OP{V}_t[k] \; \OP{\rho}(t) +
\OP{V}_t[k] \; d\OP{\rho}(t) + d\OP{V}_t[k] \; d\OP{\rho}(t)
\ee
with $d\OP{\rho}(t)$ from (\Ref{SDM}), and $\OP{V}_t[k]$ obeys an equation
similar to (\Ref{dVout}). Taking the trace over the bath,  $\Tr_B
\{ \ldots \}$, as in the derivation of the reduced density matrix (\Ref{DM}), we obtain
the following equation for the characteristic reduced density operator
\be \Label{chi}
\frac{d}{dt} {\chi}_{t}[k] = \left( {\cal L}  +
(e^{i k(t)}-1) {\cal J} \right) {\chi}_{t}[k] \quad (\chi_0[k] =
\rho) \; ,
\ee
with ${\cal L}$ the Liouville operator from Eq.~(\Ref{DM}), and $\cal J$ the
``recycling operator'' (\Ref{recyclingop}).
Note that for $k(t)=0$ this characteristic density operator coincides with the
system density operator, $\chi_t[k=0]=\rho(t)$, and Eq.~(\Ref{chi}) reduces to
the master equation (\Ref{DM}).

We solve Eq.~(\Ref{chi}) by iteration
\bea
\chi_t[k] =  \Sc (t,0) \rho
 + \sum_{m=1}^{\infty} && \int_{0}^{t} dt_m \int_{0}^{t_m} dt_{m-1}
\ldots  \int_{0}^{t_2} dt_1
\exp \left( i \sum_{r=1}^{m} k(t_r) \right) \nonumber  \\
&& \Sc (t,t_m) \Jc (t_m) \Sc (t_m,t_{m-1}) \ldots \Sc
(t_2,t_1) \Jc (t_1) \Sc (t_1,0) {\rho} \nn
\eea
with propagators
\bea
&&\Sc (t,t_0) \rho  =  \Ueff (t,t_0) \rho \Ueff^\dagger (t,t_0)  \quad (t \ge
t_0), %
\\
&&{\dot U}_{\rm eff} (t,t_0)  = - i \Heff (t) \Ueff (t,t_0)  \quad
(\Ueff (t_0,t_0) =\one) \; .
\eea
Taking a trace over the system degrees of freedom, and comparing with
(\Ref{defCF}) we can express the EPDs in terms of system averages
\bea
&&P_0^{t}(0|{\rho}) = \Tr_S \{ \Sc (t,0) {\rho
} \}, %
\\
&&p_{0}^{t}(t_1,t_2,\ldots,t_m | {\rho})= \nonumber \\
&&\quad\Tr_S  \{ \Sc (t,t_m) \Jc (t_m) \Sc (t_m,t_{m-1}) \ldots \Sc
(t_2,t_1) \Jc (t_1) \Sc (t_1,0) {\rho} \} \; . \nn
\eea
The structure of these expressions agrees with what is expected from continuous
measurement theory according to Srinivas and Davies \Cite{SD81}.

\subsubsection{Generalization to many channels}
As a reference for the following sections, we give the generalization of these
equations to $\Nd$ channels:
\bea
&&P_{t_0=0}^t (0 | \rho) = \Tr_S
\left\{ \Sc (t,0) \rho \right\} \, ,\\ 
&&p_{t_0=0}^{t}(j_1,t_1;\ldots;j_m,t_m|\rho)=\nn \\ 
&&\quad\Tr_S  \left\{
  \Sc (t,t_m) \Jc_{j_m} (t_m) \Sc (t_m,t_{m-1}) \ldots \Sc (t_2,t_1) \Jc_{j_1}
(t_1) \Sc (t_1,0) \rho
\right\} \nonumber
\eea
where $t_0=0 < t_1 < \ldots \ t_m \le t$ and $j_k=1,\ldots,\Nd $.

\subsubsection{Examples of the  use of the characteristic functional}
We conclude our discussion with a few examples for the application of
the characteristic functional:
The mean intensity of the outgoing field is
\bea
\AV{\OP{I}(t)} = - i \frac{\delta}{\delta k(t)} \Phi_{T}[k] |_{k=0}
=\Tr_S \{ \Jc  {\rho}\targ{t} \} \quad \mbox{(T > t)} .
\eea
The intensity correlation function can be written as the sum of
a normally ordered contribution plus a shot noise term
\bea
\AV{\OP{I}(t) \OP{I}(t')} &=&
\AV{:\OP{I}(t) \OP{I}(t'):} +  \delta (t-t') \, \AV{\OP{I}(t)} \nn%
\\
& =& (- i)^2 \frac{\delta^2}{\delta k(t) \, \delta k(t')} \Phi_{T}[k]
|_{k=0} %
\\
&=& \Tr \{  \Jc (t) {\cal T}(t,t') \Jc (t') {\rho}\targ{t'}  \} + \delta(t-t')
\Tr_S \{ \Jc (t)  {\rho}\targ{t} \} \; . \nn
\eea
The normally ordered $m$-th order intensity correlation function is
\be
\AV{ : \OP{I} (t_1) \ldots \OP{I} (t_m) :}
=
\Tr_S  \{ \Jc  (t_m) \Tc (t_m,t_{m-1}) \ldots \Tc
(t_2,t_1) \Jc (t_1) \Tc (t_1,0) {\rho} \}
\ee
with $\Tc (t,t_0) = \exp {\cal L} (t-t_0)$ the time evolution operator
for the density matrix according to the master equation.

\subsection{Conditional Dynamics and A Posteriori States}
\Label{aposteriori}

If during a continuous measurement a certain trajectory of a measured
observable is registered, the state of the system conditional to this
information is called a {\em conditional state} or {\em a posteriori
state} denoted by $\rho _c (t)$, and the corresponding dynamics is
referred to as {\em conditional dynamics}
\Cite{Carmichael,BB91,GPZ92,WisemanThesis}. In the present case of
photodetection (counting processes) we assume that the counters have
registered in the time interval $(t_0=0,t]$ the sequence
$j_1,t_1;\ldots;j_n,t_n$ ($t_1 < \ldots < t_n$). To derive an
expression for $\rho_c (t)$ Barchielli and Belavkin \Cite{BB91} consider
\be
P(0,(t,t+{\bar t}\,]|\rho) =
\frac{
p_0^{t+\bar t}\, (j_1,t_1;\ldots; j_n,t_n | \rho) }{ p_0^{t}
(j_1,t_1;\ldots; j_n,t_n | \rho) }
\ee
which is the conditional probability of observing no count in $(t,t+\bar t\,]$
given $\rho$
at time $t_0=0$ and having observed the sequence $j_1,t_1;\ldots;
j_n,t_n$.
Obviously one can write $P(0,(t,t+{\bar t}]|\rho) = P_t^{t+\bar
t}(0|\rho_c(t))$ with
\be \Label{CPap}
\rho_c(t) = \Sc (t,t_n) \Jc _{j_n} (t_n) \Sc (t_n,t_{n-1})
\ldots \Sc (t_2,t_1) \Jc _{j_1}(t_1) \Sc (t_1,0) \rho / \Tr_S  \{ \ldots\}
\ee
where $\Tr_S  \{ \ldots \}$ is a normalization factor to ensure $\Tr_S
\rho_c (t) =1$. Similar arguments can be given for other EPDs. Thus we
identify (\Ref{CPap}) with the {\em a posteriori state}.

\vspace{2ex}
\noindent Discussion \Cite{Carmichael,BB91,GPZ92,WisemanThesis}:
\begin{Discussion}
\item The interpretation of Eq.~(\Ref{CPap}) is as follows:
When no count is registered the system evolution is given by
$\Sc(t,t_0)$ and the state of the system between two counts is
\be \Label{propdm}
\rho_c (t) = \Sc (t,t_r) \rho_c (t_r) /\Tr_S \{\Sc (t,t_r) \rho_c (t_r)  \}
\quad \mbox{($t \ge t_r$)}
\ee
where $t_r$ is the time of the last count, and $\rho_c(t_r)$ is the
state of the system just after this count.

When a count $j$ at time $t=t_r$ is registered the action on the system is
given by the operator $\Jc _j$, and the state of the system
immediately after is
\be \Label{qjdm}
\rho_c(t_r + dt) = \Jc_j (t_r) \rho_c (t_r) / \Tr_S \{ \Jc_j (t_r) \rho_c (t_r)
\}\; .
\ee
We will call this a ``quantum jump.''
\item
When the system is initially in a pure state described by the state
vector $\psi$, $\rho=\ket{\psi} \bra{\psi}$, it will remain in a
pure state $\psi_c (t)$, $\rho_c(t) =\ket{\psi_c(t)}\bra{\psi_c(t)}$.
The time evolution between the counts is
\be \Label{propwf}
\pct (t) = \Ueff (t,t_r) \pct (t_r), \quad \psi_c (t) =
\pct (t)/\norm{\pct (t)}
\ee
i.e.
\be \Label{propwff}
 \frac{d}{dt} \pct (t) = -i \Heff \pct (t)
\ee
where ${\pct (t)}$ and ${\psi_c (t)}$ denotes the unnormalized and
normalized wave functions, respectively, and a count $j,t_r$ is
associated with the quantum jump
\be \Label{qjwf}
\pct (t_r +dt) = \lambda_j \cop{j} \pct (t_r) \; .
\ee
The complex numbers $\lambda_j \ne 0$ (with dimension $[\mbox{\rm
time}]^{1/2}$) are arbitray and can be chosen, for example, to
renormalize the wave function $\pct (t)$ after the jump.

\item
In preparation for the wavefunction simulation to be discussed in
Sec.\Ref{WavefunctionSimulation} we note:
The mean number of counts of type $j$ in the time interval $(t,t\!+\!dt]$
conditional upon the observed trajectory $j_1,t_1;\ldots;j_n,t_n$ is
\be \Label{meannumberofcounts} \Label{meancountrate}
p_t^{t\!+\!dt}(j,t|\rho_c(t)) \: dt =
\mycase{
\Tr_S  \{ \Jc_j(t) \rho_c(t) \}\: dt}{
}{
\g{j} \norm{\cop{j} \psi_c(t) }^2 \: dt}{
} \; ,
\ee
and
the joint probability that no jump occured in the time interval
$(0,t]$, and a jump of type $j$ occured in $(t,t\!+\!dt]$ given $\rho$ at time
$t=0$ is
\be \Label{nextcount}
p_t^{t\!+\!dt} (j,t|
\; \rho_c(t)
\; ) \times P_0^t (0| \rho) =
\mycase{
\Tr_S  \{ \Jc_j (t) \Sc (t,0) \rho \} \: dt
}{}{
\g{j} \norm{  \cop{j}\,  \Ueff (t,0) \psi  }^2 \: dt
}{} \; .
\ee
with $\rho_c(t)
= { \Sc (t,0) \rho / \Tr_S \{\Sc (t,0) \rho  \} }$.
\end{Discussion}

\subsection{Stochastic Schr\"odinger Equation and Wave Function Simulation for
Counting Processes}
\Label{sec:SSE}
\subsubsection{Stochastic Schr\"odinger Equation}
We can combine the time evolution for the (unnormalized) conditional
wave function $\pct (t)$ according to Eqs.\ (\ref{propwf},\ref{propwff})
into a single (c--number) Stochastic
Schr\"odinger Equation (SSE) \Cite{BB91,WisemanThesis}. A typical trajectory
$\N{j}{t}$, where $\N{j}{t}$ is the number of counts up to time $t$,
$j=1,\ldots,\Nd $ is a sequence of step functions, such that
$\N{j}{t}$ increases by one if there is a count of type $j$ and
$\N{j}{t}$ is constant otherwise. Therefore, the \Ito\/ differential
\be
\dN{j}{t} \equiv \N{j}{t\!+\!dt} - \N{j}{t} =
\mycase{1}{\mbox{if count $j$ in $(t,t\!+\!dt]$}}{0}{\mbox{else}}
\ee
fulfills $[\dN{j}{t}]^2 = \dN{j}{t}$. Moreover the probability of more than one
count in an interval $dt$ vanishes faster than $dt$. We thus have the
\Ito\/ table
\be \Label{Itotablecount}
\dN{j}{t}\; \dN{k}{t} = \delta_{jk} \;\dN{j}{t}, \quad \dN{j}{t}\: dt = 0 
\ee
and obtain a
{\em Stochastic Schr\"odinger Equation for the unnormalized wave function $\pct
(t)$}
\bea \Label{SSEu}
d \pct(t) & \equiv & \pct (t\!+\!dt) - \pct(t) %
\\  & = &
\left\{
-i \Heff \: dt + \sum_{j=1}^\Nd \left(
\lambda_j \cop{j} -\one \right) \dN{j}{t}
\right\} \pct (t) \;  \nn
\eea
with $\lambda_j \ne 0$. According to (\Ref{meannumberofcounts}) the
mean number of counts of type $j$ in the interval $(t,t\!+\!dt]$
conditional to the trajectory $j_1,t_1;\ldots;j_n,t_n$ (indicated by
the subscript $c$) is
\be \Label{DictateJumps}
\AV{\dN{j}{t}}_c = \g{j} \norm{\cop{j} \psi_c(t)}^2 \: dt \; .
\ee

The SSE (\Ref{SSEu}) can be converted to an equation for the
normalized $\psi_c (t) = \pct(t)/\norm{\pct (t)}$. As an illustration for Ito
calculus we give details in Appendix \Ref{ItoExcercise}. The result is
\bea \Label{SSEn}
d \psi_c(t) & \equiv & \psi_c (t\!+\!dt) - \psi_c(t)  \nn
\\  & = &
\Big\{
-i H \,dt -\half\sum_{j=1}^\Nd \g{j} \left(\cop{j}^\dagger \cop{j} - \AV{
\cop{j}^\dagger \cop{j}}_c \right) \; dt  \\
&& +
\sum_{j=1}^\Nd
\left(
 \nofrac{\cop{j}}{\sqrt{\AV{\cop{j}^\dagger \cop{j}}_c}} -1\right) \dN{j}{t}
\Big\} \psi_c (t) \nn \; .
\eea
where
\be \Label{expect}
\AV{\ldots}_c= \bra{\psi_c(t)}\ldots\ket{\psi_c(t)} \; .
\ee
 Eq.~(\Ref{SSEn}) is a {\em nonlinear} Schr\"odinger equation.  (Note:
Eq.~(\Ref{SSEn}) is not unique up to a phase factor: $\psi_c (t)
\rightarrow e^{i \phi (t)}\psi_c (t)$ where $\phi(t)$ is solution of a
stochastic differential equation $d\phi (t) = a(t) \: dt +
\sum b_j(t) \dN{j}{t}$ with arbitrary $a(t)$ and $b_j (t)$.)

\subsubsection{Equation of motion for the Stochastic Density Matrix}
Finally, we derive an equation for the {\em Stochastic Density Matrix} $\rho_c
(t) =\ket{\psi_c(t)} \bra{\psi_c(t)}$,
\bea \Label{SDMn}
d \rho_c(t) & \equiv & \rho_c (t\!+\!dt) - \rho_c(t) %
\\  & = & {\cal L}
\rho_c(t)\: dt +
\sum_{j=1}^\Nd
\left(
\frac{\Jc_j \rho_c(t)}{\Tr_S \{ \Jc_j \rho_c(t)  \}} -\rho_c(t) \right)
\left(
\dN{j}{t} - \right.\nn \\
&&\left. \Tr_S \{ \Jc_j \rho_c(t) \} \: dt
\right) \nn \; .
\eea

\noindent Discussion:
{\em The Master Equation} (\Ref{masterequation}) is (re)derived by taking the
stochastic mean of Eq.~(\Ref{SDMn}).  We first take the
mean for the time step $t \rightarrow t\!+\!dt$ conditional upon a given
trajectory.  All quantities on the right hand side of Eq.~(\Ref{SDMn})
depend only on the past (they are nonanticipating or adapted
functions), but the {\em a posteriori} mean value of $\AVc{\dN{j}{t}}$ is
given by (\Ref{DictateJumps}), and thus the last term vanishes.  Then
we take mean values also on the past, and obtain the master
equation (\Ref{masterequation}): $\dot \rho (t) = \Lc \rho (t)$. Thus, if the
results of a measurement
are not read, i.e. no selection is made, the state of the system at
time $t$ will be $\rho (t) = \AVS{\rho_c (t)}$, and $\rho (t)$ is the
{\em a priori} state for the case of continuous measurement.

From our construction it is obvious that the statistics of the jumps
$N_j (t)$ is identical to the statistics of
photon counts discussed in Sec.~\ref{sec:PhotonCounting}. However, as a
consistency check and for pedagogical purposes, we will rederive the
counting statistics in the framework of the characteristic functional
for the counting process $N_j(t)$ and show that it is identical to the characteristic functional derived in Sec.~\Ref{sec:CF}.
We confine ourselves to $\Nd=1$. In the present context the charcteristic functional for the counting process $N_1(t)\equiv N(t)$ is defined as
\be \Label{CF2}
\Phi_t[k] = \AVS{V_t[k]} \mbox{with} \quad V_t[k]:=\exp \left( i
\int_0^t k(s) \, \dN{}{s}  \right) \;
\ee
which should be compared with the definition (\Ref{CFV}) for the output {\em operator} $\Lambda^\out \targ{t}$.
The quantity  $V_t[k]$ satisfies the c--number SDE
\be \Label{dV}
dV_t[k] = V_t[k] \, \left(  e^{i k(t)} -1 \right) \dN{}{t} \; .
\ee
In analogy to Eq.~(\Ref{chidef}) we introduce
\be \Label{chidef2}
\Phi_t[k] = \Tr_S \chi_t[k] \quad \mbox{with} \quad \chi_t[k] :=
\AVS{V_t[k]\,  \rho_c(t)} \; .
\ee
According to Ito calculus
\be \Label{Itocount}
d [ V_t[k] \; \rho_c(t)] =  d V_t[k] \; \rho_c(t) + V_t[k] \; d\rho_c(t)] +
dV_t[k] \; d\rho_c(t)
\ee
where $dV_t[k]$ and $d\rho_c(t)$ obey Eqs.~(\Ref{dV}) and (\Ref{SDMn}),
respectively.  Repeating the arguments given in the derivation of the
density matrix equation from the stochastic density matrix equation
given above, we take a stochastic average $\AVS{\ldots}$ of the equation  for $V_t[k] \; \rho_c(t)$ and find
that $\chi_t[k] \equiv \AVS{V_t[k] \; \rho_c(t)} $ (Eq.~(\Ref{chidef2})) satisfies the equation
of motion and initial condition (\Ref{chi}) given in Sec.~\Ref{sec:PhotonCounting}. This completes the proof that the characteristic
functional defined in Eq.~(\Ref{CF2}) is identical to the
characteristic functional of Sec.
\Ref{sec:PhotonCounting}.

A direct derivation of the c--number SSE (\Ref{SSEn}) from the QSSE
(\Ref{QSSEvac}) has been given by Gardiner {\em et al.~} \Cite{GPZ92}. Goetsch
and Graham \Cite{G94} have given a derivation by working with a
eigenrepresentation of the set of commuting operators
$d\Lambda\targ{t}$.


\subsubsection{Wave Function Simulation}
\Label{WavefunctionSimulation}
The conditional dynamics defined by Eqs.~(\Ref{propwf},\Ref{qjwf})
together with (\Ref{nextcount}) suggests a {\em wave function
simulation} of the reduced system {\em density matrix} $\rho\targ{t}$ as
follows
\Cite{Carmichael,B90,BB91,GPZ92,WisemanThesis,%
G94,DCM92,M93,DRZ92,DPZG92,%
N95,Bel,CBA92,Marte,Dum94,CBSDM94,NKS95,CM95,SGK95,GK94a,GK94b,C93}:

\vspace{2ex}

\begin{Discussion}
\item[Step A:] We choose a system wave function
$\pct (t_0)$, initially normalized $\norm{\pct(t_0)}=1$, and set the counter
$n$ for the
number of quantum jumps equal to zero, $n=0$.
\item[Step B:] We propagate the (unnormalized) conditional wavefunction $\pct$
according to Eq.~(\Ref{propwf})
and simulate the time $t$ and type of count $j$ of the {\em next
quantum jump} according to the conditional density (\Ref{nextcount}),
\be
\tilde c (j,t) =\norm{c_j \pct (t) }^2 
\quad   \mbox{($t
\ge t_n,\norm{ \pct (t_n) }=1 $)}.
\ee

\begin{Remark} Note: One possible way to determine $t$ and $j$ is to proceed
in two steps.  First, we find a decay time $t$ according to the {\em
delay function} $\tilde c(t)= \sum_{j} \tilde c(j,t)$.  This is
conveniently done by drawing a random number $0 \leq r \leq 1$ from a
uniform distribution and monitoring the norm of $\pct(t)$ until
\bea\Label{eq:2.15}
\int^t_{t_n} dt'
\tilde c (t')  & \equiv & 1 - \norm{ \pct (t) }^2 %
\\
&=& r \; \epsilon\; [0,1]\; \quad \mbox{($t\ge t_n$; $\norm{\pct(t_n)}=1$).}
\nn
\eea
Second, the type of count $j$ is determined from the conditional
density $\tilde c(j|t) = \tilde c(j,t)/\sum_{k} \tilde c (k,t)$ for
the given time $t$.
\end{Remark}

Incrementing $n \rightarrow n+1$ we identify $t_n \equiv t$ and $j_n
\equiv j$ with the decay time and type of count, respectively.  The
wave function after the quantum jump is given by Eq.~(\Ref{qjwf}). We
renormalize this wavefunction and continue integrating (\Ref{propwf})
up to the next jump time, i.e. return to the beginning of Step B.

\item[Step C:] An approximation for the system density matrix is obtained by
repeating these simulations in steps A and B to obtain
\begin{equation}\Label{eq:2.17} \Label{DMSimulation}
\rho(t) =
\AVS{ \ket{\psi_c(t)}\bra{\psi_c(t)}   }
\end{equation}
where $\AVS{\ldots}$ denotes an average over the different
realizations of system wave functions.
\end{Discussion}

\vspace{2ex}
\noindent Remarks:
\begin{Discussion}
\item
A {\it simulation} of the quantum master equation in terms of {\it
system wave functions} can replace the {\em solution} of the master
equation for the {\it density matrix}, and an important feature of the
wave function approach, first emphasized by Dalibard, Castin and
M{\o}lmer \Cite{DCM92}, is that one has only to deal
with a wave function of dimension $N$, as opposed to working with the
density matrix and its  $N^2$ elements. Thus, simulations can provide
solutions when a direct solution of the master equation is
impractical because of the large dimension of the system space.
Convergence of the simulation vs. a direct solution of the master
equation in terms of ``global'' and ``local'' observables are given in
Ref.~\Cite{M93}.

In many problems one is interested in a steady state density matrix. In this
case it is often convenient to replace the ensemble averages by the time
average of a  single trajectory.

A simulation has the further advantage that it allows the calculations
to be performed on a distributed system of networked computers
\Cite{Marte}, with a corresponding gain in computational power.  In
view of the statistical independence of the different wavefunction
realizations, parallelization of the algorithm is trivial.

\item
The $\Nd$--channel master equation
(\Ref{masterequation}) is form invariant under the transformation
\be \Label{trans1}
\gs{j} \cop{j} \rightarrow \sum_{k=1}^\Nd U_{jk} \gs{k} \cop{k} \quad
\mbox{where } (\sum_{j=1}^\Nd U_{jk}U_{lk}^*=\delta_{jl})
\ee
(that is with $U$  a unitary matrix). The decomposition of
Eq.~(\Ref{masterequation}) to form quantum
trajectories $\pct (t)$ is thus not unique. Different sets of jumps
operators $\{ \cop{j} \}$ not only lead to a different physical
interpretation of trajectories, but an appropriate choice of $\cop{j}$
may be crucial for the formulation of an efficient simulation method
for estimating the ensemble distribution.  An example will be given in
Sec.~\Ref{sec:OpticalMolasses} \Cite{Holland95} (see also \Cite{M93}).

\item Any master equation conforming to the requirements for  $\rho$ is of the
{
\em Lindblad form} \Cite{QN}
\be
{\dot \rho} = -i [H,\rho]
+ \sum_{i,j=1}^{\Nd} \gamma_{ij} \left(2\cop{i}\rho \cop{j}^\dagger
- \cop{j}^\dagger \cop{i} \rho - \rho\cop{j}^\dagger \cop{i} \right) \; .
\ee
This can be brought into the form (\Ref{masterequation}) by
diagonalizing the Hermitian matrix $\gamma_{ij}$ by a unitary
transformation $V_{i\gamma}$,
\be
\gamma_{ij} = \sum_{\gamma=1}^{\Nd} V_{i\gamma} \kappa_{\gamma} (V_{\gamma
i})^
\dagger \quad \mbox{($i,j,=1,\ldots,\Nd $)}
\ee
with eigenvalues $\kappa_\gamma \ge 0$, and defining
\be
\cop{\gamma} = \sqrt{\kappa_\gamma} \sum_{i=1}^{\Nd} \cop{i} V_{i \gamma} \; .
\ee
Applications of this procedure to the master equation for squeezed
noise can be found in Ref.~\Cite{DPZG92}.
\end{Discussion}

\subsection{Simulation of Stationary Two-Time Correlation Functions}
\Label{sec:correlation}

We are often interested in system correlation functions $\AV{A(t)
B({t_0})}$ and their Fourier transform (spectra) (see section
\Ref{sec:OpticalMolasses}). According to the quantum fluctuation
regression theorem the system correlation function can be written as
\be \Label{corrfct}
\AV{A(t) B({t_0})} = \Tr_S \{ A e^{{\cal L } (t-{t_0})} B \rho ({t_0}) \} \quad
(t \ge {t_0}) \; .
\ee
This correlation function can be obtained by solving the density
matrix equation for $\rho (t)$ and an equation for the first order
response $\rho ^{(+)} (t)$
\bea
&& {\dot \rho } (t)  =  {\cal L} \rho (t), \Label{response1}
\\
&& {\dot \rho}^{(+)}(t) = {\cal L} \rho ^{(+)} (t) + B f(t) \rho (t) \; .
\Label{response2}
\eea
By formally integrating Eq.~(\Ref{response2}),
\be
\rho ^{(+)} (t) = \int^t  dt_1 e^{{\cal L} (t-t_1)} B f(t_1) \rho (t_1)
\ee
and choosing a $\delta$--function $f(t)=\delta (t-{t_0})$ we have
\be
\Tr_S  \{ A\rho ^{(+)} (t)  \}= \AV{A (t) B({t_0})} \quad (t \ge {t_0}) \; .
\ee

We will show that these correlation functions can be obtained by
solving (simulating) the set of stochastic Schr\"odinger equations
\Cite{GPZ92,DPZG92,Marte}
\bea
&& d \pct(t) = \left[ -i \Heff \: dt + \sum_{j=1}^d (\lambda_j\cop{j}-\one)\dN{j}{t}
\right] \pct (t) \Label{hom}%
\\
&& d \pct^{(+)}(t) = \left[ -i \Heff \: dt +
\sum_{j=1}^d (\lambda_j\cop{j}-\one)\dN{j}{t} \right] \pct ^{(+)}(t) + f(t) B \pct
(t) dt
 \Label{inhom} \nonumber
\eea
with quantum jumps dictated by $\pct (t)$ according to
$\AV{\dN{j}{t}}_c = \norm{\cop{j} \psi_c(t)}^2 \: dt$.  For a
$\delta$--kick $\pct^{(+)} (t)$ obeys the same Schr\"odinger equation
as $\pct (t)$ (Eq.~(\Ref{hom})), where the inhomogeneous term in
(\Ref{inhom}) translates into the initial condition
\be \Label{correlationinitialcondition}
\pct^{(+)}({t_0}+dt) = B \pct ({t_0}) \; .
\ee

It is straightforward to show that the stochastic averages
\bea
&&\rho (t) = \AVS{ \ket{\pct (t)} \bra{\pct (t)}/ \norm{\pct (t)}^2 } %
\\
&&\rho ^{(+)} (t) = \AVS{ \ket{\pct ^{(+)} (t)} \bra{\pct (t)}/ \norm{\pct
(t)}^2 }
\eea
(both normalized with respect to $\pct (t)$) obey Eqs.~(\Ref{response1},
\Ref{response2}), and thus
\be
\AV{A(t) B({t_0})}=
\big\langle \;
 \AVS{ \bra{\pct (t)} A \ket{ \pct ^{(+)} (t)} / \norm {\pct (t)}^2}
\; \big\rangle _{{t_0}}
\ee
where $\AV{\ldots} _{{t_0}}$ indicates an average over radmonly selected initial ``kick''--times $t_0$.
A physical interpretation of this procedure is summarized in the
discussion following Eq.~(\Ref{OM:correlation}).  The above derivation
emphasizes the simulation of a correlation function. We can simulate a
spectrum directly when instead of the $\delta$-kick we use a function $f(t)
\propto \exp
\left( -i \nu t \right)$ \Cite{GPZ92,DPZG92,TC92,Marte}.
For another approach to simulate correlation functions we refer to
\Cite{M93}.


\subsection{Diffusion Processes and Homodyne Detection}
\Label{sec:homodyne}

Le us consider  homodyne
detection as shown in Fig.~\Ref{fig:ch}b
\Cite{Carmichael,B90,WisemanThesis}. In the simplest configuration, the
output field of the system $b^\out (t)$ is sent through a beam
splitter of transmittance close to one. The other port of the beam
splitter is a strong coherent field which serves as a local
oscillator. For a single output channel, the transmitted field is then
represented by $b^\out (t)' = b^\out (t) + \gamma =
\gs{} \cop{}+\gamma+ b(t)$ .  Assuming a real field $\gamma$ the operator
for photon counting of the outgoing field is
\bea
d\Lambda^\out (t)'/ dt &\equiv & b^\out(t){^\dagger}' b^\out (t)'
\\
& =& \gamma^2 +\gamma \left[\gs{}  c+ \gs{} c^\dagger + b (t) + b(t)^\dagger
\right] +
\left[ \gs{} c^\dagger + b(t)^\dagger \right] \left[ \gs{} c + b(t) \right] \;
. \nn
\eea
The ideal limit of homodyne detection is for infinite amplitude of the
local oscillator. Physically, in this limit the count rate of the
photodetectors will go to infinity, but we can be define an operator
for the homodyne current $\OP{I}^{\rm hom} (t)$ by subtracting the
local oscillator contribution:
\bea
\OP{I}^{\rm hom} (t) &=&  \lim_{\gamma
\rightarrow \infty} \frac{d\Lambda^\out \targ{t}'-\gamma^2 \dt}{\gamma \dt} \;
.
\eea
Defining quadrature operators for the in and out field,
\be
 d\OP{X (t)}:= dB (t) + dB^\dagger (t), \quad d\OP{Y} (t) := -i dB (t) + i dB
 (t)^\dagger
\ee
and for the
system dipole
\be
 x:=c+c^\dagger, \quad y := -i c + i c^\dagger \; ,
\ee
we can write
\bea \Label{QSDEXout}
 d\OP{X}^\out (t) &=& \OP{I}^{\rm hom} (t)  \, dt  %
\\
 &=& \gs{} x(t) \, dt +  d\OP{X} (t) \;  \nonumber
\eea
where the second line is the QSDE for the $X$--quadrature of the
out--field which follows from Eq.~(\Ref{dBout}).  This equation shows
that homodyne detection with $\gamma$ real corresponds to a
measurement of the compatible observables $b^\out (t) + b^\out (t)
^\dagger = d\OP{X}^\out (t) /dt$ ($t \ge 0$), the quadrature
components of the field.

The complete statistics of the homodyne current is obtained from a
characteristic functional \Cite{B90}
\bea \Label{CFVdiffusion}
\Phi_t[k] &= &
\AV{ \OPVout_t [k] } \quad \mbox{with} \quad \OPVout_t[k] :=
\exp \left( i \int_0^{t}   k(s) \, d\OP{X}^\out \targ{s} \right).
\eea
The characteristic operator is determined by the QSDE
\footnote{
We note the connection between Eq.~(\ref{QSDEVdiffusion}) and normal ordering relation
\benn
\exp \left( i \int_0^{t}   k(s) \, d\OP{X}^\out \targ{s} \right)
=
:\exp \left( i \int_0^{t}   k(s) \, d\OP{X}^\out \targ{s}
-\half \int_0^{t}   k(s)^2 \; ds
 \right) :
\eenn
}
\bea \Label{QSDEVdiffusion}
&& d\OPVout_{t}[k] = \OPVout_{t}[k] \, \left( i k(t)\, d\OP{X}^\out (t) -
\half k(t)^2 \, dt \right) \quad (\OPVout_0[k]=\one)
\eea
together with Eq.~(\Ref{QSDEXout}) for $d \OP{X}^\out (t)$.  Following the
derivations of Sec.
\Ref{PhotonCounting}
we define a characteristic density operator $\chi_t [k]$ which gives the
characteristic functional $\Phi_t [k]$ (\Ref{CFVdiffusion}) according to
(\Ref{chidef}). We obtain the equation
\be \Label{chidiffusion}
\frac{d}{dt} \chi_{t}[k] =  {\cal L}  \chi_t [k] -\half k(t)^2 \chi_t[k] + i
k(t)
\left(  \gs{} c \chi_t[k] + \chi_t[k] \gs{} c^\dagger \right) \quad
\mbox{where ($\chi_0[k]=\rho(0)$)}
\ee
corresponding to a Gaussian diffusive measurement \Cite{BB91}. An expression for the mean homodyne current and current - current correlation function will be given below in Eqs.~(\Ref{homc}) and (\Ref{homc2}), respectively.

\subsection{Stochastic Schr\"odinger Equation and Wave Function Simulation for
Diffusion Processes}

We now turn to a c--number stochastic description and the derivation
of a c--number stochastic Schr\"odinger equation for diffusion
processes.  Such an  equation was first derived by Carmichael
\Cite{Carmichael} (see also \Cite{WisemanThesis}) from an analysis of
homodyne detection, and independently in a more formal context by
Barchielli and Belavkin \Cite{BB91}.

From (\Ref{meancountrate}) the
mean rate of photon
counts is
\bea \Label{meancountratehomodyne}
\AVc{\dN{}{t}} &=& \Tr_S  \{ (\gs{} \cop{}^\dagger + \gamma)(\gs{} \cop{} +
\gamma) \rho_c (t)  \} \: dt %
\\  \nn
&=& \Tr_S  \{ (\gamma^2 + \gamma \gs{} x + \g{} \cop{}^\dagger \cop{}) \rho_c
(t)\}
\: dt \; ,
\eea
and registration of a count in homodyne detection is associated with a
quantum jump $\pct \leftarrow (c+\gamma) \pct$ of the system wave
function.
Furthermore,  we note that the master equation is invariant under the
transformation
\bea \Label{trans2}
\cop{} &\rightarrow& \cop{} +\gamma %
\\  H & \rightarrow & H -  i \half
(\gamma^* \gs{} \cop{} - \gamma \gs{} \cop{}^
\dagger)
\eea
which involves a displacement of the jump operators by a complex number
$\gamma$. With these replacments we obtain
from the SSE  (\Ref{SSEn})
\bea \Label{SSEnhomodyne}
d \psi_c(t) & = &
\left(
-i H -
\half \left(
\g{} \cop{}^\dagger \cop{} + 2 \gamma \gs{} \cop{}- \g{} \AVc{\cop{}^\dagger 
\cop{}} - \gamma
\gs{} \AVc{x} \right)
\right) \: dt \;  \psi_c (t)
\\&+&  
\left( 
 \frac{\gs{} \cop{}+\gamma}{\sqrt{\AV{(\gs{} \cop{}^\dagger+\gamma) (\gs{} 
\cop{} +
\gamma)}_c}} -1\right) \dN{}{t}
\;  \psi_c (t) \nn
\eea

We are again interested in the ideal limit when the local oscillator
amplitude $\gamma$ goes to infinity. As pointed out before, in the
limit that $\gamma$ is much larger than $\cop{}$, the count rate
(\Ref{meancountratehomodyne}) consists  of a large constant term
(the counting rate form the local oscillator), a term proportional to
the $x$ quadrature of the system dipole, and a small term, the direct
count rate from the system.  Subtracting from the counting rate the
contribution of the local oscillator we write
\be
\dN{}{t} = \gamma ^2 \: dt + \gamma \; dX (t)
\ee 
which defines the stochastic process $X (t)$. For $\gamma
\rightarrow \infty$ the stochastic process $X (t)$ has the
Gaussian properties
\bea \Label{propertyY}
\mbox{(i)} \quad && \AVc{{d{X{(t)}}}} = \AVc{x}\dt  \Label{Y1} %
\\  
\mbox{(ii)} \quad && [dX(t)]^2 =  \dt, \quad dX (t) \, dt =0 \Label{Y2} %
\\ 
\mbox{$\Longrightarrow$}\quad && dX (t) - \AVc{dX(t)} = dW (t)
\eea
 where $dW (t)$ is a Wiener increment $[dW (t)]^2 = dt$,
$\AVc{dW(t)}=0$.  Thus the jumps (\Ref{SSEnhomodyne}) are replaced by
a diffusive evolution.  The derivation of (\Ref{propertyY}) is straightforward
\Cite{BB91}: Eq.~(\Ref{Y1}) follows from
Eq.~(\Ref{meancountratehomodyne}); Eq.~(\Ref{Y2}) is obtained from the
\Ito\ table (\Ref{Itotablecount}),
$
[dX (t)]^2
= \dN{}{t}/ \gamma^2 =  \dt + {dX (t)} / \gamma
$.
Physically we interpret
\bea
I^{\rm hom} (t) &=& \frac{dX(t)}{dt} \equiv \lim_{\gamma
\rightarrow \infty} \frac{\dN{}{t}-\gamma^2 \dt}{\gamma \dt}%
\\  &
=& \gs{} \AVc{x} (t)+ \xi (t) \Label{homodynecurrent} \nn
\eea
as a stochastic homodyne current with $\xi (t) = dW (t) / dt$ white noise.

Taking this limit in Eq.~(\Ref{SSEnhomodyne}), we obtain a SSE for diffusive 
processes
\bea \Label{SSEdiffusive}
d \psi_c(t) & = &
\Big\{
\left(
-i H -
\half (
\g{} \cop{}^\dagger \cop{} - \g{} \AVc{x} \cop{} + \frac{1}{4}\g{} \AVc{x}^2 )
\right) \: dt  \\ &&+ 
 \gs{} \left( 
\cop{} - \AVc{x/2} \right) dW(t)
\Big\} \psi_c (t) \; . \nn
\eea
Eq.~(\Ref{SSEdiffusive}) is not unique; in particular we can make a
phase transformation $\psi_c (t) \rightarrow e^{i\phi (t)}\psi_c (t)$ which
allows us to rewrite this equation as \Cite{BB91}
\bea \Label{SSEdiffusive2}
d \psi_c(t) & = &
\Big\{
\left(
-i H -
\half \g{} \cop{}^\dagger \cop{} + \g{} \AVc{\cop{}^\dagger} \cop{} +
\half
\g{} | \AVc{\cop{}}|^2
\right) \: dt \\ &&+
\gs{}  \left(
\cop{} - \AVc{\cop{}} \right) dW(t)
\Big\} \psi_c (t) \; , \nn
\eea
and there are also versions with complex noises  \Cite{G84,GP92,DGHP95}.

The unnormalized version of the equation has been given by Carmichael \Cite{Carmichael},
and more recently by Wiseman \Cite{WisemanThesis,G94}
\be \Label{SSECarmichael} d \pct (t) =
-i H - \half \g{} \cop{}^\dagger \cop{} + I^{\rm hom}(t) \;
\gs{} \cop{} ] \dt \; \pct (t) \;
 \ee
which demonstrates clearly how the
state is conditioned on the measured photocurrent $I^{\rm hom}(t)$.

A stochastic density matrix equation for diffusive processes is directly
obtained from these equations,
\be \Label{DMdiffusive}
d \rho_c (t) = \Lc \rho_c  \, dt +
\left[\gs{} \left( c-\AVc{c} \right)\rho_c + \rho_c  \gs{} \left( c^\dagger-
\AVc{c^\dagger} \right) \right]\,  dW(t)
\ee
and taking stochastic averages we see - as expected - that the a
priori dynamics satisfies the master equation (\Ref{DM}).

As shown by Goetsch et al.~ \Cite{G94} the above c--number SSE for diffusive
processes  can be derived directly from the QSSE (\Ref{QSSEvac}). The idea is
to replace $\dB{t}^\dagger\, c
\ket{\Psi\targ{t}} \rightarrow (\dB{t} + \dB{t}^\dagger)\, c
\ket{\Psi\targ{t}}\equiv d\OP{X}\targ{t}^\dagger\, c
\ket{\Psi\targ{t}}$ in Eq.~(\Ref{QSSEvac}) and to subsequently project on an
eigenbasis of the set of commuting operators $d\OP{X}$. Generalization of the
SSEs to a squeezed bath have been given in \Cite{WisemanThesis,G94}.

The stochastic homodyne current $I^{\rm hom} (t)$
(\Ref{homodynecurrent}) has the same statistical properties as the
operator version $\OP{I}^{\rm hom} (t)$ as becomes evident by defining a
characteristic functional \Cite{BB91}
\be \Label{CFdiffusion2}
\Phi_t[k] = \AVS{V_t[k]} \mbox{with} \quad V_t[k]:=\exp \left( i
\int_0^t k(s) \, dX\targ{s}  \right)
\ee
where $V_t[k]$ satisfies the SDE
\be \Label{dVhom}
dV_t[k] = V_t[k] \, \left( i k(t) \, dX\targ{t} - \half k(t)^2 \, dt  \right)
\; .
\ee
Following (\Ref{chidef2}) we again define a ``density matrix'' $\chi _t
[k]$, and use Ito calculus (\Ref{Itocount}) with the above
equation for $V_t [k]$ and the stochastic density matrix equation for
diffusive processes (\Ref{DMdiffusive}) to find that $\chi_t[k]$ obeys an
equation and initial conditions identical to (\Ref{chidiffusion}). Thus the
characteristic functionals $\Phi _t [k]$ defined in Eqs.~(\Ref{CFVdiffusion})
and (\Ref{CFdiffusion2}), respectively, are identical.

As an example, we readily derive the mean homodyne current
\be \Label{homc}
\AV{\OP{I}^{\rm hom} (t)} \equiv
\AVS{I^{\rm hom} (t)} =
- i \frac{\delta}{\delta k(t)} \Phi_{T}[k] |_{k=0}
=
 \Tr _S x \rho \targ{t} \nn
\ee
and the stationary homodyne correlation function whose Fourier transform gives
the spectrum of squeezing \Cite{Carmichael}
\bea \Label{homc2}
\AV{\OP{I}^{\rm hom} (t_1) \OP{I}^{\rm hom} (t_2)} &\equiv&
\AVS{I^{\rm hom} (t_1) I^{\rm hom} (t_2)} \nn \\
& =& (- i)^2 \frac{\delta^2}{\delta k(t) \, \delta k(t')} \Phi_{T}[k]
|_{k=0}
\\
&=&\g{} \Tr_S \left[ x e^{\Lc (t_2-t_1)}\left( c \rho + \rho c^\dagger
\right) \right] + \delta (t_2-t_1) \nn
\eea
where the last term is the shot noise contribution.

As a final comment we note that a completely different interpretation
of the stochastic Schr\"odinger equations of the type
(\Ref{SSEdiffusive}) have been proposed in Refs. \Cite{G84,GP92,DGHP95} in
connection with {\em dynamical theories of wave
function reduction}. The assumption is that the reduction of the
wavefunction associated with a measurement is a stochastic
process and an equation of the type (\Ref{SSEdiffusive2}) is postulated.

There have been numerous applications of the diffusive SSE. A
beautiful discussion of squeezing in a parametric amplifier in terms
of trajectories can be found in Carmichael's book \Cite{Carmichael}.
Illustrations for the decay of Schr\"odinger cats have been given in
\Cite{G94,GK94a,GK94b,K95LesHouches}. A theory of quantum feedback has been
developed by Milburn and Wiseman \Cite{WisemanThesis}.

\section{Application and Illustrations}
\Label{sec:application}

\mbox{During the last years we have seen numerous, mostly numerical}
applications of the SSE
\Cite{Carmichael,B90,BB91,GPZ92,WisemanThesis,%
G94,DCM92,M93,DRZ92,DPZG92,N95,%
Bel,CBA92,Marte,Dum94,CBSDM94,NKS95,CM95,SGK95,GK94a,GK94b,C93}. Roughly
speaking, these
calculations can be divided into {\em illustrations} where single
trajectories illustrate certain features of the physics one wishes to
discuss, and as a {\em numerical simulation method} for situations where a
direct solution of the master equation is not feasible. Below we
discuss a few examples from our recent work. Our first example is state
preparation in ion traps through quantum jumps; the second example is a
discussion of simulations in laser cooling, and we conclude with
remarks on possible application to the problem of decoherence in quantum
computing.

\subsection{State Preparation by Observation of Quantum Jumps in an Ion Trap}

Ions traps provide a tool to store and observe laser cooled {\em
single ions} for an essentially unlimited time. Experiments with
single trapped ions thus represent a {\em realization of
continuous observation of a single quantum system} in the context of
quantum optics. For a review on ion traps and application in quantum optics we
refer to  \Cite{Walther94,Blatt95}.

\subsubsection{Quantum Jumps in Three-Level Atoms}

The probably best-known example is the problem of ``observation of
quantum jumps in three level system'' \Cite{qjexp} (for a review and references
see \Cite{BZ}).  The system of
interest involves a double--resonance scheme  where two excited states $
\ket{e}$ and
$\ket{r}$ are connected to a common lower level $\ket{g}$ via a strong
and weak transition, respectively. The fluorescent photons from the
strong transition are observed. However, an excitation of the weak
transition where the electron is temporarily shelved in the metastable
state $\ket{r}$ will cause the fluorescence from the strong transition
to be turned off.  It is, therefore, possible to monitor the quantum
jumps of the weak transition via emission windows in the (macroscopic)
signal provided by the fluorescence of the strong transition.
Experimental observation of this effect has been reported in Refs.
\Cite{qjexp}, and various theoretical treatments of this effect have
been published. From a theoretical point of view, the problem is to
study the photon statistics of the photons emitted on the strong line
of the three--level atom. A treatment based on
continuous measurement theory was given by Zoller et al.~ \Cite{ZMW87} and
Barchielli \Cite{B87} (see also \Cite{CD86}):
in Ref. \Cite{ZMW87} the probability density for the ``emission of the next
photon on the strong line'' (delay function) was calculated, and for the first
time a simulation of the a posteriori dynamics was
described to illustrate the conditional dynamics of a single
continuously monitored quantum system.

The master equation for a three--level atom
 has the form
\be \Label{masterequation3}
{\dot \rho} = - i (\Heff \rho - \rho \Heff^\dagger) + \Jc _{\rm s}
\rho + \Jc _{\rm w} \rho \quad (\equiv \Lc \rho)
\ee
where $\Jc_{\rm s} \rho : = \ket{g} \bra{g} \, \Gamma_s \rho _{ee}$
and $\Jc_{\rm w} \rho : = \ket{g} \bra{g} \, \Gamma_w \rho _{rr}$ are
the ``recycling operators'' for the atomic electron
on the strong and weak transition, respectively. The probability
density for the emission of a photon on the strong line at time $t$,
given the previous photon on the strong transition was emitted at
$t_r$ is, according to Eq.~(\Ref{meannumberofcounts}),
\bea \Label{delayfunction3}
{\tilde c}(s,t|s,t_n) &=&
p_{t_n}^{t_n\!+\!dt}(s,t|\: \rho_c (t_n) \: )/dt
\nn %
\\
&=& \Tr_S \{ \Jc _{\rm s} e^{(\Lc-\Jc_{\rm s})(t-t_n)} \Jc _{\rm s} \rho \}
\over \Tr_S \{ \Jc_{\rm s} \rho \}
\eea
where $\rho_c (t_n) = \ket{g} \bra{g}$ is the density matrix at time
$t_n$ after emission of a photon on the strong line at time $t_n$. We
note that this conditional state is independent of the previous
history of photon emissions; this is due to the fact that a quantum
jump always prepares the atom in the ground state $\ket{g}$. In
addition, we have summed over an arbitrary number of possible
transitions on the weak line which replaces $\Sc \targ{t,t_n}
\rightarrow \exp {(\Lc-\Jc_{\rm s})(t-t_n)}$ in Eq.~(\Ref{qjdm}),
$\rho$ is the steady state density matrix.  It is obvious from
(\Ref{masterequation3})\Cite{ZMW87} Eq.~(\Ref{delayfunction3}) is a
sum of exponentials with different decay constants, reflecting the
different time scales for the excitation and decay on the strong and
metastable transition, which - when these distributions are simulated
- manifest themselves in the periods of brightness and darkness in the
fluorescence on the strong transition. In
particular, the density matrix conditional to observing an emission
window on the strong line when the last s-photon was emitted at time
$t_n$ is from Eq.~(\Ref{propdm})
\be \Label{propdm3}
\rho_c (t) = e^{(\Lc - \Jc_{\rm s})(t-t_r)} \rho  /\Tr_S \{ \ldots  \} {\;
\rightarrow \;} \ket{r} \bra{r} \quad \mbox{($\Gamma_s (t-t_n) \gg 1$)}\; ,
\ee
i.e. {\em observation of a window} in a single trajectory of counts
corresponds to a {\em preparation of the electron in the metastable state}
$\ket{r}$ (``shelving of the electron'').

\subsubsection{Fock states in the Jaynes-Cummings model}

The principle of state preparation by quantum jumps can be extended to
more complicated configurations.  As a second example, we discuss the
{\em preparation of Fock states in a Jaynes-Cummings Model (JCM) based
on the observation of quantum jumps} \Cite{Cirac93} (see also
\Cite{Toschek95}). The JCM describes a harmonic oscillator strongly
coupled to a single two-level atomic transition. The Hamiltonian is
\begin{equation}\Label{eq:JCM}
H = \hbar \omega_f a^\dagger a + \frac{1}{2}\hbar\omega_{eg}
\sigma_z + \hbar g (\sigma_+a + a^\dagger\sigma_-) ,
\end{equation}
where $a^\dagger$ and $a$ are creation and annihilation operators for
a mode of the radiation field with frequency $\omega_f$, and
$\sigma_{\pm ,z}$ are the Pauli spin matrices describing a two--level
atom with transition frequency $\omega_{eg}$. The last term in
(\Ref{eq:JCM}) describes the coupling of the field mode to the atom
with coupling strength $g$. Dissipation can be included in this model
by coupling the field mode and atom to independent heatbaths and using
a master equation formulation, in which one introduces damping rates
$\kappa$ and $\Gamma$ for the field mode and atom respectively.  Of
particular interest is the strong--coupling regime in which $g>\kappa$ when
quantum effects
in the coupled oscillator--spin system are most pronounced. Here the
spectroscopy of the system is best described in terms of transitions
between the ``dressed states,'' or eigenstates, of the Hamiltonian
$H$.  The ground state is $|0,g\rangle =|0\rangle |g\rangle$ and, on
resonance ($\omega_0=\omega_f$), the excited dressed states are
$|n,\pm\rangle =(|n-1\rangle |e\rangle \pm |n\rangle |g\rangle )
/\sqrt{2}$ ($n=1,2,\ldots$), where $|g\rangle$ and $|e\rangle$ are the
bare atomic ground and excited states respectively, and $|n\rangle$
are  Fock states of the field mode with excitation number
$n=0,1,2,\ldots$.  The eigenenergies corresponding to the excited
states are $E_{n,\pm}=\hbar[\omega_0(n-1/2)\pm g\sqrt{n}]$, which show
AC--Stark splitting proportional to $\sqrt{n}$.  The two lowest
transitions $|1,\pm\rangle\rightarrow |0,g\rangle$ give rise to a
doublet structure, the ``vacuum'' Rabi splitting
\Cite{RH95LesHouches,K95LesHouches}.

Experimental realization of the JCM with dissipation has been
demonstrated in the field of cavity QED, both in the microwave and
optical domain \Cite{K95LesHouches,RH95LesHouches}.
\comment{
 Observation of quantum revivals and non--classical
photon statistics with Rydberg atoms and microwave cavities have been
reported in \Cite{MicrowaveQED}, and in the optical regime vacuum Rabi
splitting has been observed in \Cite{RH95LesHouches,K95LesHouches}.  In the
Rydberg atom
experiments radiation damping is negligible and the strong coupling
limit is reached by means of weakly--damped microwave cavities,
whereas in optical experiments radiation damping is significant but
strong coupling is achieved via high finesse cavities and very small
cavity mode volumes. }

An alternative realization of the JCM is a {\it single} trapped ion
constrained to move in a harmonic oscillator potential and undergoing
laser cooling at the node of a standing light wave. Under conditions
in which the vibrational amplitude of the ion is much less than the
wavelength of the light (Lamb--Dicke limit), this problem is
mathematically equivalent to the JCM with negligible damping of the
oscillator (i.e. $\kappa =0$).  In this configuration, preparation of
the Fock state corresponds to preparation of a non--classical state of
motion of the trapped ion with fixed energy $n\hbar\omega$, whereas in
cavity QED the Fock state corresponds to a non--classical state of
light with no intensity fluctuations and undetermined phase.

As discussed in detail in Refs. \Cite{Cirac93} the master equation for
single two--level ion $\{ \ket{g}, \ket{e}\}$ trapped in a harmonic
potential and located at the node of a standing light wave has the form
\begin{eqnarray} \Label{masterequationJCM}
\frac{d}{dt}\rho &=& -i\left[ \nu a^\dagger a +
(-\frac{\Delta}{2})\sigma_z
- \frac{\Omega}{2}\eta (\sigma_++\sigma_-)(a+a^\dagger),\rho
\right] \nonumber%
\\
& &+ \frac{\Gamma}{2} (2\sigma_-\rho\sigma_+ - \sigma_+\sigma_-\rho -
\rho\sigma_+\sigma_-).
\end{eqnarray}
where the oscillator now refers to the ion motion with frequency $\nu$
(phonons), $\Delta$ is the detuning of the laser from atomic
resonance, and the atom - oscillator coupling is proportional to the
Rabi frequency $\Omega$ times the Lamb--Dicke parameter $\eta$. Eq.
(\Ref{masterequationJCM}) is valid to lowest order in $\eta = \pi a_0
/ \lambda \ll 1$ (Lamb--Dicke limit) with $a_0$ the size of the ground
state wave function of the trap and $\lambda$ the wavelength of the
light.  Making the associations $-\Delta\leftrightarrow\omega_0$,
$\nu\leftrightarrow\omega_f$, and $g\leftrightarrow\eta\Omega/2$, we
have a clear analogy with the damped JCM. The additional terms
proportional to $\sigma_+a^\dagger$ and $\sigma_-a$ are usually
dropped in the optical regime on the basis of the RWA. In our
instance, such an approximation requires that $\nu ,\Delta \gg\Omega
,\Gamma ,|\nu -\Delta |$, which is consistent with the sideband
cooling limit implied above.  A significant feature of the JCM
produced by this configuration is that dissipation in the system is
entirely due to damping of the two--level transition, while the
oscillator is undamped (i.e. $\kappa =0$).  Further, the effective
coupling constant $\eta\Omega/2$ depends on the Rabi frequency
and, in contrast to CQED, can be adjusted experimentally to satisfy
the strong coupling condition $\Gamma < \eta\Omega/2$.

The spectroscopic level scheme of the Jaynes--Cummings ladder
 of the strongly coupled ion + trap system is shown in
Fig.~\Ref{fig:Fock1}.  One possible approach to the observation of this level
structure is a measurement of the probe field absorption spectrum
to a third atomic level $|r\rangle$, very weakly coupled to the
otherwise strongly coupled $|g\rangle$-$|e\rangle$ transition, analogous to
level schemes used
for the observation of quantum jumps discussed above.
We note that as a consequence of the unequal spacing of the energy
levels in the JCM, the probe laser is only resonant with the
transition frequency between a {\it single} pair of levels, and thus will only
excite the system to a
single state $|n,r\rangle$.

This ability to selectively excite a particular transition, together
with the state reduction associated with the observation of quantum
jumps, offers the intriguing possibility of generating number states
of the quantized trap motion. This follows from the fact that the
probe laser exciting transitions to the states $|n,r\rangle$ interacts
only with the atomic ground state contribution to the particular
dressed state $|n,\pm\rangle$ being excited.  Given that we are able
to
distinguish spectroscopically between the different maxima
characterizing the absorption spectrum (so that we can identify the
dressed state being excited), observation of an emission window in the presence of a weak coupling to the state $|n,r\rangle$ will tell us with certainty that
the vibrational state of the ion is $|n\rangle$ (i.e. this is the
vibrational state occurring with $|g\rangle$ in the dressed state
$|n,\pm\rangle$) and that we have {\em produced a Fock state of the
quantized trap motion}.

An obvious consequence of having the freedom to choose which
transition is excited is the ability to choose the Fock--state that is
to be produced.  As an example, Fig.~\Ref{fig:Fock4}(a) shows the simulation of
an
experiment with a trapped three-level ion and the calculated random
telegraph signal due to the probe excitation. Shelving of the electron in  the
state $|2,r\rangle$ are indicated when emission windows appear in the
fluorescence of the strongly coupled atom--trap system. Thus a Fock
state of the trap motion with $n=2$ is prepared during these dark
periods. Fig.~\Ref{fig:Fock4}(a) shows the number of photons as observed in a
real
experimental situation where the integration time constant is long
compared with the decay time of the strongly coupled system.  To
highlight the internal dynamics, we evaluate the temporal evolution of
the mean quantum number $\langle n\rangle$ and the corresponding
entropy of the system for a different set of parameters and over a
timescale in which system relaxation is clearly visible. This is shown
in Fig.~\Ref{fig:Fock4}(b) and Fig.~\Ref{fig:Fock4}(c) which demonstrate that
after a quantum jump
to the state $|r\rangle$ the system's entropy is suddenly reduced to
zero (indicating the Fock state). After the ion returns to its
strongly coupled states $|n,\pm\rangle$ the mean value $\langle
n\rangle$ approaches its thermal equilibrium value (here set by the
mean photon number $N$ characterizing the broadband thermal noise
field) and the entropy increases to its steady state value.
\begin{figure}
\begin{center}\
\psfig{file=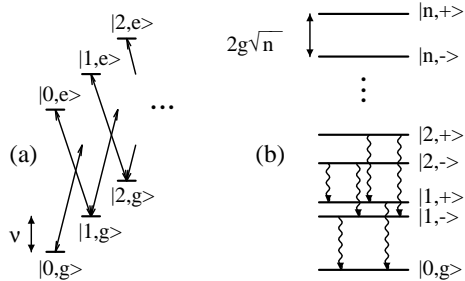,width=3in,angle=90}
\end{center}
\caption{Level scheme of ion--trap system. 
(a) Bare states $|n,g\rangle$,$|n,e\rangle$:
The laser--atom detuning is 
$\omega_L-\omega_0=-\nu$.  
Arrows pointing to the left
indicate the resonant laser coupling with 
strength $\eta \Omega$;
arrows pointing to the right correspond 
to non--rotating--wave terms. There
are no $|n,g\rangle \; - \; |n,e\rangle$ 
couplings since the
ion is at the node of the standing wave. 
(b) Dressed states $|n,\pm\rangle$ resulting from
the laser coupling. {\protect\Cite{Cirac93}} }
\Label{fig:Fock1}
\end{figure}

\subsection{Wave Function Simulations of Laser Cooling: Applications to
Quantized Optical Molasses}
\Label{sec:OpticalMolasses}

One of the prime examples of wave function
simulations in recent years is application to laser cooling of neutral atoms
(for a review see \Cite{MolmerCastin95}).  Examples are description of
quantized optical molasses in one-, two- and three-dimensional configurations
and fluorescence and probe transmission spectra
\Cite{DRZ92,M93,CBA92,Marte,Dum94,CBSDM94,NKS95,CM95}.
Here we give a brief summary of our work on the spectrum of resonance
fluorescence from quantized optical molasses, in particular the comparison
between theory and
experiment on the fluorescence spectrum as measured by Jessen {\em et al.~}
\Cite{Jessen} (see also \Cite{Verkerk,HemmerichHansch}), and more recent work
on a fully quantum mechanical description of diffusion in optical molasses
employing wave function simulations \Cite{Holland95}.
\begin{figure}
\begin{center}\
\psfig{file=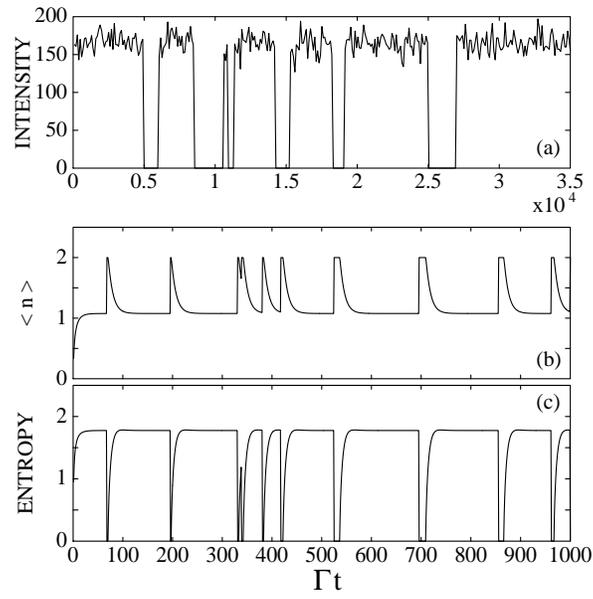,width=3in}
\end{center}
\caption{Simulation of quantum jumps as
a function of time. The probe
laser is tuned to $|2,-\rangle \rightarrow |2,r\rangle$.
(a)
Fluorescence intensity on the $|e\rangle$ - $|g\rangle$
transition
($\Omega_{\rm w}=2\times 10^{-5}
\Gamma_{\rm s}$, $\Gamma_{\rm w}= 10^{-7} \Gamma_{\rm s}$).
(b) Time evolution of
$\AV{n}$, and (c) the entropy,
$S=-\Tr \{ \rho \ln (\rho) \}$, of the system
($\Omega_{\rm w}=0.1 \Gamma_{\rm s}$) {\protect\Cite{Cirac93}}. }
\Label{fig:Fock4}
\end{figure}

\subsubsection{Quantized Atomic Motion in Optical Molasses}

Laser cooling is typically accomplished in optical molasses, employing
a configuration of counterpropagating laser beams.  The role of the
laser is twofold: it provides a damping mechanism, and leads to the
formation of optical potentials (the ac Stark shift of the atomic
ground states) \Cite{Wallis}.
 The physical picture underlying laser cooling and spectroscopy of
one--dimensional (1D) molasses is illustrated in
Figs.~\Ref{fig:mol},~\Ref{fig:MolResFlu1} and \Ref{fig:MolResFlu2} for an
angular momentum $J_g=1/2$ to $J_e=3/2$ transition. Figure \Ref{fig:MolResFlu1}
shows the
atomic configuration for this model \cite{CastinDalibard,DalibardCohen}. We
consider two counterpropagating linearly polarized laser beams with
orthogonal polarizations, so that the positive frequency part of the
electric field has a position dependent polarization
\be \Label{OM:EF} \mbox{$\bf E$}^{(+)}_{\rm cl}(z,t) = {\cal
E} (\sin(kz) \mbox{$\bf e$}_{+1} +
\cos(kz)\mbox{$\bf e$}_{-1}) e^{-i \omega t}
  \ee where $\omega$ is the frequency, $k=2 \pi/\lambda$ the wave
vector with $\lambda$ the wavelength of the laser light, and
$\mbox{$\bf e$}_{\pm 1}$ are spherical unit vectors.  For red laser
detunings $\Delta=\omega -\omega_{eg}$ and low laser intensities, i.e.
small saturation parameter $s={\Omega^2/2 / (\Delta^2 + \Gamma^2/4)}
<< 1$ with $\Omega$ the Rabi frequency and $\Gamma$ the spontaneous
decay width, the Stark shifts of the two $\ket{g_{-}}$ and
$\ket{g_{+}}$ ground states will form an alternating pattern of
optical bipotentials $U_{\pm} (z)$. This is shown in the upper part of
Fig.~\Ref{fig:MolResFlu2}: due to the large Clebsch--Gordan coefficients for the
outer
transitions (see Fig.~\Ref{fig:MolResFlu1}), minima will occur for the state
$\ket{g_{\pm}}$ at positions with pure $\sigma^{\pm}$--light. In
addition, spontaneous emission causes transitions between these
potentials via optical pumping processes.  In the semiclassical
picture of Sisyphus--cooling \cite{DalibardCohen} one considers an atom moving
on one of these potential curves, say $U_{-}(z)$. Transitions to the
other potential $U_{+}(z)$ then occur preferentially from the tops of
$U_{-}(z)$ down to the valleys of $U_{+}(z)$, so that on the average
the atomic motion is damped.  Quantum mechanically, laser cooling can
be understood as optical pumping between the quantized energy levels
(band structure) in the limit of level separations $\hbar
\omega_{\rm osc}$ much larger than the optical pumping rate $\gamma_0 =
s \Gamma$ (which implies large laser detunings) \Cite{CastinDalibard}.
In the time domain this condition corresponds to a situation in which
an atomic center-of-mass wave packet in the potential undergoes many
oscillations with frequency $\omega_{\rm osc}$ before an optical
pumping process occurs \Cite{Marte}. An example of a band structure
is shown in the upper part of Fig.~\Ref{fig:MolResFlu2}. As a result of laser cooling the atom
will occupy the lowest energy levels, and will thus be strongly
localized. The lower part of \Ref{fig:MolResFlu2} shows the corresponding
localization of the atom in minima of the $U_{\pm } (z)$ potentials.
Transitions between the vibrational states will manifest themselves as
sidebands (Raman transitions due to optical pumping) in the atomic
absorption and emission spectra \Cite{Jessen,Verkerk,HemmerichHansch}.

The basis of a theoretical discussion of laser cooling is the solution of the
Generalized Optical Bloch Equations for the atomic density matrix
comprising both the internal and external (center--of--mass) degrees
of freedom \Cite{Wallis}.
The corresponding master equation for the density matrix $\rho$ of the
ground state manifold and 1D motion in the $z$--direction is
\Cite{Marte,CastinDalibard} ($\hbar=1$)
\begin{eqnarray} \Label{OM:GOBE}
\dot \rho = &-&i({H_{\rm eff}}(\hat{z})\rho - \rho {H_{\rm
eff}}(\hat{z})^{\dagger}) \nonumber %
\\  &+& \gamma_0
\sum_\sigma\int_{-1}^{+ 1} du \; N_\sigma(u)
B_\sigma(\hat{z}) e^{-ik u\hat{z}}\rho B_\sigma(\hat{z})^{\dagger}
e^{ik u\hat{z}}.
\end{eqnarray}
$N_\sigma(u)$ is the angular distribution of spontaneous photons
emitted in the direction $u = \cos ( \vec{k}_s,\vec{e}_z )$ with
polarization $\sigma=0,\pm 1$, and $\gamma_0= s \Gamma /2 $ is the
photon scattering rate.  The first two terms on the right-hand side of
the master equation involve the non-Hermitian atomic Hamiltonian
\begin{equation} \Label{OM:Heff} {H_{\rm eff}}(\hat{z})=
\frac{\hat{p}^2}{2M} - (U_0+i \frac{1}{2} \gamma_0) \sum_\sigma
B_\sigma(\hat{z})^{\dagger} B_\sigma(\hat{z})
\end{equation}
describing the motion of the atomic wavepacket with kinetic energy
$\hat{p}^2/2M$ in a multicomponent optical potential with depths
determined by $U_0={s |\Delta| / 2}$. The real part of the potential
in (\Ref{OM:Heff}) gives rise to quantized energy levels (band
structure), while the imaginary part describes a loss rate due to
optical pumping.  The operators $B_\sigma (z)$ correspond to Raman
transitions between the ground state levels by absorption of a laser
photon and subsequent emission of a spontaneous photon with
polarization $\sigma$, see Ref.\Cite{Marte}.
The master equation (\Ref{OM:GOBE}) is of the Lindblad form with an infinite
number of channels (the $u$ integration over the $z$--projection of the emitted
photon wave vector in Eq.~(\Ref{OM:GOBE}))

We are interested in the spectrum of resonance fluorescence, emitted
along the $z$-axis, with frequency $\omega'$ and polarization
$\sigma$. This spectrum is proportional to the Fourier transform of the
stationary atomic dipole correlation function \Cite{Marte}
\begin{equation}
c_{\sigma}(t-t_0)=\langle B_{\sigma}(t)^{\dagger}\; e^{ik'_z\hat{z}(t)}\;
 B_{\sigma}(t_0)\; e^{-ik'_z\hat{z}(t_0)}\rangle \; .
\end{equation}
According to Section \Ref{sec:correlation} and Ref.~
\Cite{Marte} we
have \begin{equation} \Label{OM:correlation} c_{\sigma}(t-t_0)=\Tr_S
\left\{B^{\dagger}_{\sigma} \; e^{ik_z
\hat{z}}\;\rho^{(+)}(t)\right\}\quad (t \geq t_0) \;\;\;;
\end{equation}
where the first order perturbed density operator $\rho^{(+)}(t)$
obeys the same master equation as the density matrix but with a
different initial condition
\begin{equation} \Label{OM:initialcondition}
\rho^{(+)}(t_0)=B_{\sigma}(\hat{z})\; e^{-ik'_z
 \hat{z}} \rho(t_0)\; .
\end{equation}

Physically, this simulation procedure can be interpreted as the
following computer experiment \Cite{GPZ92,DPZG92,Marte} (for a more
formal derivation see Sec.~\Ref{sec:correlation}). We surround the
atom by an (infinite number of) unit efficiency photodetectors
covering the $4 \pi$ solid angle.  The simulated ``click'' of the
photodetectors determines the time, polarization and direction of the
emitted photon (see Fig.~\ref{fig:mol} a).  This is the simulation of the density matrix of the
laser cooled atom.  To simulate the fluorescence spectrum, we place a
Fabry-Perot in front of one of these photodetectors and ``measure''
the transmitted photon flux. This gives the spectrum as a function of
the tuning of the resonance frequency of the Fabry-Perot transmission
(see Fig.~\ref{fig:mol} b).

In the experiment by Jessen {\em et al.~} the resonance fluorescence
spectrum from Rb$^{85}$ atoms in 1D molasses is observed. The Rb atoms
are driven on the $5S_{1/2}$ $F_g=3 \rightarrow 5P_{3/2}$ $F_e=4$
transition in a lin $\perp$ lin laser configuration described above.
In this case a direct solution of the master equation (\Ref{OM:GOBE})
to calculate the autocorrelation function (\Ref{OM:correlation}) is
impractical due to the large dimensionality of the density matrix
equation which involves $N^2$ elements ($N=N_{ex}\times N_{\rm int}$
where $N_{int}$ is the number of internal, and $N_{ex}$ the number of
(discretized) external degrees of freedom).  For $^{85}$Rb we have $N
= 448$ on a Fourier grid with 64 points corresponding to momenta up to
$\pm 32 \hbar k$ \Cite{Marte}; $10000$ wavefunction realizations are
needed for convergence.
\begin{figure}
\begin{center}\
\psfig{file=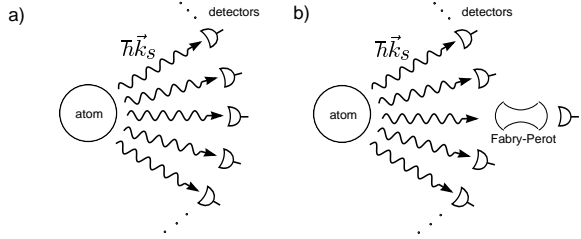,width=3in,angle=90}
\end{center}
\caption{
(a) The ``experiment'' simulated in the Monte Carlo wave function
description of optical molasses: the atom is surrounded by photo
detectors corresponding to an angle resolved detection of the emitted
photon $\hbar \vec{k}_s$.  (b) Simulation of the
fluorescence spectrum of the emitted light: the atom is ``observed'' with photodetectors and the spectrum is obtained by filtering one output channel with  a Fabry-Perot interferometer {\protect\Cite{Marte}}.  }
\Label{fig:mol}
\end{figure}
Adapting the formalism of Sec.~\Ref{WavefunctionSimulation} a
simulation of the master equation (\Ref{OM:GOBE}) consists of
propagation of an atomic wave function $\pct (t)$ with the
non--Hermitian (damped) atomic Hamiltonian (\Ref{OM:Heff}) interrupted
at random times by wave function collapses
\begin{equation} \Label{OM:qj} {{ {\pct (t\!+\!dt)}  }}
= e^{ik'_z\hat{z}}B_\sigma(\hat{z}){{ {\pct (t)}
}},
\end{equation}
and subsequent wave function renormalization.  The Schr\"odinger
equation for ${{ {\pct (t)}  }}$ describes the time evolution
of the atomic wave packet in the periodic optical potential, and its
coupling to the laser driven internal atomic dynamics. The times of
the ``quantum jumps'' are selected according to the delay function
\begin{equation}
\Label{OM:delayfunction} \tilde c (u',\sigma,t) = \norm{
\sqrt{\gamma_0 \;N_\sigma(u')  } \; B_\sigma(\hat{z})\pct (t)   }^2
\end{equation}
which gives the probability for emitting a spontaneous photon at time
$t$, with momentum $k'_z=k u'$ along the z-axis and polarization $\sigma$.
The ``quantum jump'' (\Ref{OM:qj}) corresponds to an optical pumping
process between the atomic ground states, including the associated
momentum transfer to the atom.  Averaging over these wave function
realizations gives the density matrix  according to Eq.~(\Ref{DMSimulation}).

The dipole correlation function (\Ref{OM:correlation})
can be simulated following the approach outlined in section
\Ref{sec:correlation}
\Cite{GPZ92,DPZG92,Marte}.  The perturbed density matrix
$\rho^{(+)}(t)$ in Eq.~(\Ref{OM:correlation}) can be interpreted as a
first order response to a ``delta-kick'' at time $t=t_0$, represented
by the initial condition
(\Ref{OM:initialcondition},\Ref{correlationinitialcondition}). A
simulation is obtained by introducing a ``perturbed'' wave function
${{ {\pct^{(+)}(t)} }}$ which obeys the Schr\"odinger equation for ${{
{\pct (t)} }}$ but now with initial condition [compare
Eq.(\Ref{OM:initialcondition},\Ref{correlationinitialcondition})]
\begin{equation} {{ {\pct^{(+)}(t=t_0)}  }} =
e^{-ik'_z \hat{z}} B_\sigma(\hat{z}) {{ {\pct(t_0)}
} },
\end{equation}
and quantum jumps of ${{
{\pct^{(+)}(t)}  } }$ dictated by the wave
function ${{ {\pct(t)}  }}$ according to the
delay function (\Ref{OM:delayfunction}).  The dipole correlation function
is
\begin{equation} c_{\sigma}(t-t_0)= \AVS{  \bra{ {\pct(t)}
} B_{\sigma}(\hat{z})^{\dagger} e^{ik'_z \hat{z}} {\ket{
{\pct^{(+)}(t)} }} / \norm{\pct(t)}^2
}
\end{equation}
where the angular brackets indicate averaging over
both quantum jumps and initial times $t_0$.

The periodicity of the atomic Hamiltonian (\Ref{OM:Heff}) in space,
and assuming an infinite periodic molasses, allows one to propagate
the atomic wave packets as {\em time dependent Bloch functions}
\begin{equation}
\pct(z,t) =
{1 \over \sqrt{2\pi} }\; e^{iq z} u_q(z,t) 
\end{equation}
with $q \in (-k,k]$ a quasi momentum in the first Brillouin zone, and
$u_q(z,t)=u_q(z+\lambda/2,t)$ a periodic multicomponent atomic wave
function.  The Hamiltonian evolution due to $\Heff$ preserves $q$,
while quantum jumps cause changes between families of Bloch functions,
$q\rightarrow q'$. In practice, we propagate the Bloch function
$u_q(z,t)$ on the unit cell of the lattice $z \in [0,\lambda)$ using a
split-operator Fast-Fourier Transform method \Cite{Marte}.

Figure \Ref{fig:RbSpectrum} compares the resonance fluorescence
spectrum for $\sigma^{+}$ polarized light obtained by simulation
(solid line) with the experimental data of Jessen et al.~\Cite{Jessen}
(crosses) for laser intensities, detunings etc.  taken directly from
the experiment with no adjustable parameter. The theoretical spectrum
of Fig.~\Ref{fig:RbSpectrum} was obtained by convolving the {\em ab
initio} spectrum with the experimental resolution. The central line is
scattering at the laser frequency while the first red and blue
sidebands correspond to Raman transitions between adjacent vibrational
bands in the optical potential.
The asymmetry of the red and blue sideband intensities reflects the
populations of the vibrational levels, and from the excellent
agreement between theory and experiment we infer that the wave
function simulation reproduces the experimental temperature of the
atoms\Cite{Jessen}.  Laser cooling accumulates atoms predominantly in
the lowest vibrational states and in the $M_g= \pm 3$ potentials. This
spatial localization of atoms -- on a scale small compared with the
laser wavelength -- suppresses optical pumping transitions between
different vibrational levels $n \ne n'$, which is responsible for a
narrowing of the lines in the optical spectrum (Lamb--Dicke
narrowing).  A broadening mechanism is present for the sidebands, due
to the anharmonicity of the optical potential. This leads to different
transition frequencies for $n \rightarrow n\pm 1$ (by approx. $ E_R /
\hbar$).  For the present parameters this anharmonicity is not
resolved even in the unconvolved spectrum.
\begin{figure}
\begin{center}\
\psfig{file=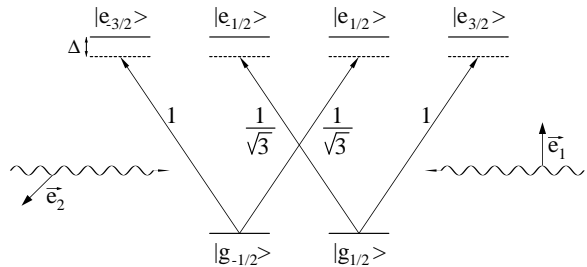,width=3in,angle=270}
\end{center}
\caption{Atomic level scheme and Clebsch-Gordan coefficients for a
$J_g=1/2$ to $J_e=3/2$ transition in a lin $\perp$ lin configuration
{\protect\Cite{Marte}}.
 }\Label{fig:MolResFlu1}
\end{figure}
\begin{figure}
\begin{center}\
\psfig{file=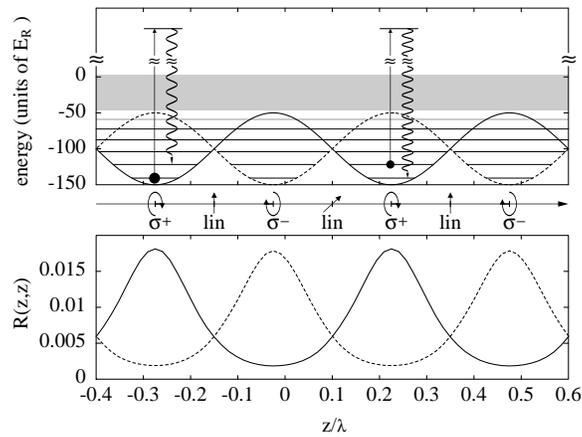,width=3in,angle=90}
\end{center}
\caption{In the upper panel the optical bipotentials
and bandstructure of the atom are plotted as a function of position $z$
for a $1/2$ to $3/2$ transition. The potential depth is $U_0=100 E_R$.
The excited states were adiabatically eliminated. We schematically
indicate the two Raman processes between the ground and excited states
which lead to the red and blue sidebands in resonance fluorescence. In
the lower panel we show the spatial distribution of the
atoms in the $\ket{g_{+}}$--state (solid line) and
$\ket{g_{-}}$--state (dashed line) for $U_0=100E_R$ and
${\gamma}_0=5/3 E_R$. The atoms are localized in the valleys of the
corresponding optical potentials {\protect\Cite{Marte}}.  }
\Label{fig:MolResFlu2}
\end{figure}
\begin{figure}
\begin{center}\
\psfig{file=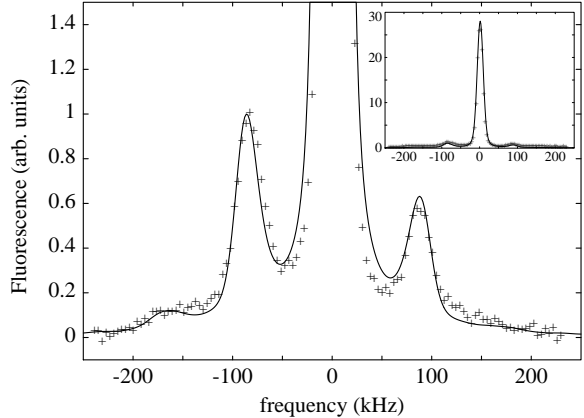,width=3in,angle=90}
\end{center}
\caption{Spectrum of resonance fluorescence as a
function of the frequency
$\nu$. The
solid line
is the theoretical spectrum {\protect\Cite{Marte}}  convolved with a Lorentzian
corresponding
to a
finite detector width of $3.8$ kHz and a Gaussian with
width $20$ kHz
(residual Doppler broadening).  Crosses are the
experimental result
of Jessen et al. {\protect\cite{Jessen}}.  The inset shows the total
spectrum.
For parameters we refer to Ref.~{\protect\Cite{Marte}}. }
\Label{fig:RbSpectrum}
\end{figure}

\subsubsection{Localization by Spontaneous Emission}
\Label{OpticalMolasses}

As outlined in the context of Eq.~(\Ref{trans1}) there is no unique
way of decomposing a given master equation (\Ref{masterequation}) to
form quantum trajectories $\pct (t)$, since the reservoir measurement
may be performed in any basis. This statement is equivalent to noting
that Eq.~(\Ref{masterequation}) is form invariant under the
substitution (\Ref{trans1}). Different sets of jump operators $\{
\cop{j} \}$ not only lead to a different physical interpretation of
trajectories, but an appropriate choice of $\cop{j}$ may be
crucial for the formulation of an efficient simulation method for
estimating the ensemble distribution \Cite{Holland95}.  We will
illustrate these ideas in the context of the quantized motion of an atom
moving in optical molasses for the master equation (\Ref{OM:GOBE}).

Simulation methods for modeling spontaneous emission according to the master
equation  (\Ref{OM:GOBE}) have assumed an {\em angle resolved detection
of the photon} as illustrated in Fig.~1(a). For a one dimensional
system with adiabatically eliminated excited state this gave the jump operators
(\Ref{OM:qj}),
\begin{equation}
 c_{u \sigma}=\sqrt{ \gamma_0 \, N_\sigma(u)} \, e^{-ik u
\hat{z}}B_\sigma(\hat{z}) \quad \mbox{($-1 \le u \le +1$, $\sigma=0,\pm
1$ )} \; . \Label{OM:direct1D}
\end{equation}
In a simulation each quantum jump gives information on the direction
of the emitted photon.  Alternatively, we could {\em observe the
fluorescence through a lens} (Fig.~\Ref{fig:loc1}(b)). The wave function simulation
in this case is equivalent to the direct simulation of a Heisenberg
microscope~\Cite{Pfau}: it is not possible to distinguish different
paths {$u_1$} and {$u_2$} (Fig.~\Ref{fig:loc1}(b)) which may be taken by the photon so the
decay operator does not generate a unique recoil. Instead, information
about the position coordinate is provided by each emission event, and
as a result {\em the wavefunction is localized in position space}.
Applying a Fourier transform to the operators in Eq.~(\Ref{OM:direct1D})
to model the action of the ideal lens gives the new decay operators
for one dimension
\begin{equation}
  c_{\nu\sigma}=\int_{-1}^1du\, \sqrt{\gamma_0 N_{\sigma}(u)}
B_\sigma (\hat z) e^{-iku(\hat x-\nu\lambda/2)} \quad \mbox{($\nu=0,\pm
1, \ldots$)} \; .
\Label{OM:localizing}
\end{equation}
For angle resolved detection, $u$ labels a continuous but bounded set
of operators, so that in the conjugate basis, $\nu$ can be any integer
and indexes an infinite set of operators at discrete points. The
integral can be evaluated for $\hat c_{\nu=0\,\sigma}$ to give a
localized function centered at the origin and the rest generated by
translation by multiples of $\pm \lambda/2$.  To prove that both sets
of jumps operators (\Ref{OM:direct1D}) and (\Ref{OM:localizing}) give
rise to the same a priori dynamics, we consider the following identity
for the recycling term in Eq.~(\Ref{OM:GOBE}),
\bea
&\Jc&_\sigma \rho \propto \int\limits_{-1}^{1} du\,
\left( \sqrt{N_\sigma(u)}B_\sigma e^{ik u \hat{z}} \right)
\rho
\left( \sqrt{N_\sigma(u)} B_\sigma e^{ik u \hat{z}} \right)^\dagger
 =  
\\
&=& \sum \limits_{\nu=-\infty}^\infty
\left( \int\limits_{-1}^{1} du\, \sqrt{N_\sigma (u) \over 2} B_\sigma e^{ik u (\hat{z}-
\nu \lambda/2)}  \right)
\rho 
\left( \int\limits_{-1}^{1} du'\, \sqrt{N_\sigma (u') \over 2} B_\sigma e^{ik u' (
\hat{z}-\nu \lambda/2)}  \right)^\dagger \nn
\eea
where we have used
\be
\delta (u-u') = \half \sum\limits_{\nu=-\infty}^\infty  e^{i \pi \nu (u-u')}
\quad \mbox{($-1 < u,u' < +1$)} \; .
\ee

As an application of the new simulation method, we consider a fully
quantum mechanical treatment of atomic diffusion in optical molasses
for the lin $\perp$ lin laser configuration described above
\Cite{Holland95}.  The calculation of quantum diffusion using the
angle resolved detection approach is difficult because the wave
function spreads out during the coherent propagation. In contrast to a
description of an infinitely extended periodic molasses~\Cite{Marte},
we have here an intrinsically non-periodic problem.  Applying the
localizing jump operators (\Ref{OM:localizing}) allows us to make use
of a greatly reduced basis set which we allocate dynamically to follow
the atom.  One representative trajectory illustrating the random walks
is shown in Fig.~\Ref{fig:loctraj}(a) where we plot the expectation
value of the spatial coordinate as a function of time. The long
periods when the position does not change appreciably correspond to
sub-barrier motion when the total energy of the atom is below the
threshold given by the maximum of the optical potential. Energy
fluctuations allow eventually the atom to overcome the potential
barrier as indicated by the dashed line in
Fig.~\Ref{fig:loctraj}(b). It may then travel over several wavelengths
until it is trapped again.  For a more complete discussion of the
spatial diffusion coefficient we refer to
Ref.~\Cite{Holland95}. Other candidates for application of this
method are cold collisions \Cite{radiativeHeating}.
\begin{figure}
\begin{center}\
\psfig{file=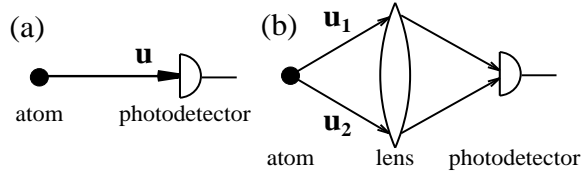,width=3in}
\end{center}
\caption{  Schematic illustration of the two different measurement
bases for the detection of spontaneous emission {\protect\Cite{Holland95}}.
}\Label{fig:loc1}
\end{figure}
\begin{figure}
\begin{center}\
\psfig{file=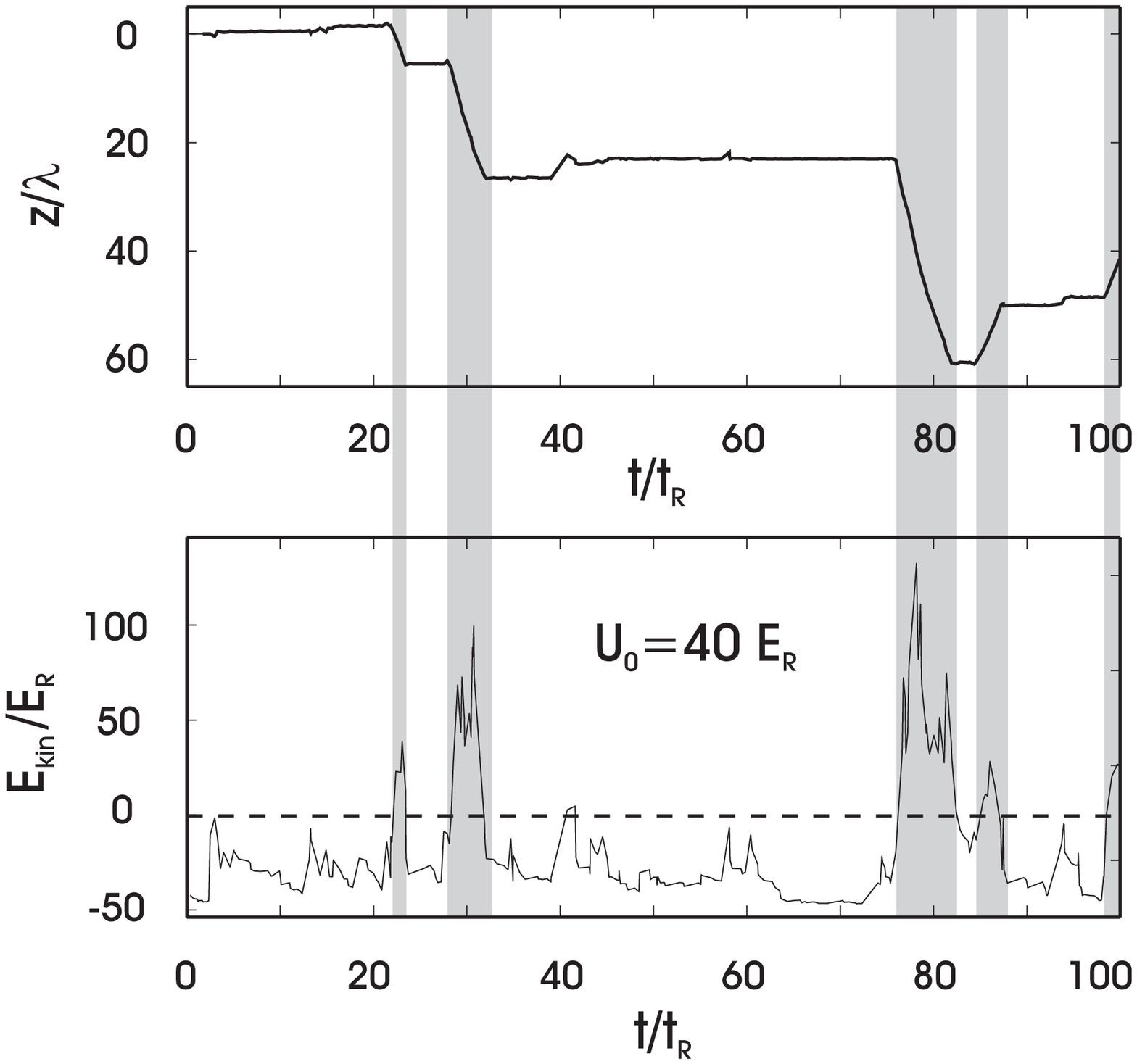,width=3in}
\end{center}
\caption{  (a) Expectation value of the spatial coordinate for a single
trajectory versus time.~(b) Corresponding (kinetic plus potential)
energy expectation value. Note the coincidences between above barrier
energies and long flight periods over many wavelengths {\protect\Cite{Holland95}}.
}\Label{fig:loctraj}
\end{figure}

\subsection{Quantum Computing, Quantum Noise and  Continuous Observation}

Quantum computers (QCs) (for a review see \Cite{Ekert95}) promise
exponential speedup for certain classes of computational problems
\Cite{Shor94} at the expense of exponential sensitivity to
noise
\Cite{Unruh95,Landauer95,Chuang95,Palma95}. Thus the analysis and
suppression of quantum noise is a fundamental aspect of any
discussion of the practicality of quantum computing.  The theory of
quantum noise and continuous measurement discussed in the preceding
sections provides the theoretical basis for such an analysis.

Computation is a process which transforms an initial state to a
final state by the application of certain rules.  These input and
output states are represented by physical objects, and computation is
the physical process which produces the final from the initial state.
Quantum
computers are physical devices which obey the laws of quantum
mechanics, in the sense that the states of the computer are state
vectors in a Hilbert space, and the processes are the unitary dynamics
generated by the Schr\"odinger equation on this Hilbert space. The
distinctive feature of QCs is that they can follow a
superposition of computational paths simultaneously and produce a
final state depending on the interference of these paths \cite{Ekert95}.
Recent results in quantum complexity theory, and the development of
some algorithms indicate that quantum computers can solve certain problems
efficiently which are considered intractable on classical Turing
machines. The most striking example is the problem of factorization of
large composite numbers into prime factors \cite{Shor94}, a problem which is the
basis of the security of many classical key cryptosystems: to factor a
number $N$ of $L$ digits on a classical computer requires a time
approximately proportional to $\exp L^{1/3}$ for the best known
algorithms, while Shor has shown that for a QC the execution time grows only
as a polynomial function $L^2$ \Cite{Shor94}.

\subsubsection{Realization of the elements of a quantum computer}
The basic elements of the QC are the quantum bits or qubits
\Cite{Ekert95}. In a classical computer a bit can be in a state $0$ or
$1$, while quantum mechanically a qubit can be in a superposition
state $\ket{\psi} = c_0
\ket{0} + c_1 \ket{1}$ with $\ket{0}$ and $\ket{1}$ two orthogonal
basis states in a two-dimensional state space ${\cal H}_2$. Quantum registers
are
defined as product states of $L$ qubits,
\be
|\underline{x}\rangle=|x_{L-1}\rangle_{L-1} \ldots
|x_{0}\rangle_{0} \in
{\cal H}_2 \otimes \ldots {\cal H}_2
\ee
with and $x_n=0,1$, and $x=\sum_{n=0}^{L-1} x_n 2^n$
the binary decomposition of the number $x$.
There are $2^L$ different states and the most general state of  a quantum
register is the superposition
\be \Label{QCregister}
|\psi\rangle =
\sum_{
\underline{x}=\{  0, 1  \}^L
} c_{\underline{x}}|\underline{x}\rangle \; .
\ee
In a QC the physical objects are state vectors $\Psi$ in the product
space (\Ref{QCregister}), and the computation is the Schr\"odinger
time evolution ${\psi _{\rm in}}\rightarrow {\psi _{\rm out }} =
\OP{U} {\psi _{\rm in}}$ with $\OP{U}$ a unitary time evolution
operator. Note that unitarity of $\OP{U}$ implies that the computation
is reversible.

\vspace{2ex}
\noindent Physical problems in realizing a QC are:
\begin{Discussion}

\item A physical realization for the qubits must be found. Examples are
internal states of atoms, or photon number states or polarization states of
photons.

\item We must be able to erase the content of the register and to prepare the
initial superposition state $\psi_{\rm in}$.

\item The unitary evolution operator $\OP{U}$ must be implemented.
 The unitary transformation can be decomposed into a sequence of
steps involving the conditional dynamics of a few qubits (quantum
gates) \cite{Ekert95}.

In particular, it has been shown
that any operation can be decomposed into
controlled--NOT gates between two qubits and rotations on a
single qubit, where a controlled--NOT is defined by
\be
\hat{C}_{12}: |\epsilon_1 \rangle  |\epsilon_2\rangle
\rightarrow  |\epsilon_1\rangle |\epsilon_1 \oplus
\epsilon_2\rangle \quad \mbox{$\epsilon_{1,2}=0,1$}
\ee
where $\oplus$
denotes addition modulo $2$.
\comment{Among the interesting properties of the controlled--NOT gate is that
it transforms product states into entangled states (Bell states) as discussed
by Ekert \Cite{Ekert95}.}
The question then is to find physical processes (i.e. a
Hamiltonian) which implement these quantum gates.

\item To read the state of the  register $\psi_{\rm out}$ at the end
of the computation we must implement and perform a von Neumann state
measurement. These measurements are destructive, i.e. irreversible.

\item In practice the central obstacle to building a QC is the fragility of
macroscopic
superpositions with respect to decoherence due to coupling of the QC to the
environment, and the inaccuracy inherent to state measurements
\cite{Unruh95,Landauer95,Chuang95,Palma95}.  Schemes and experimental
realizations must be
found which minimize these sources of decoherence.

\end{Discussion}
\vspace{2ex}
Realization of quantum gates has been discussed in Refs.~\cite
{RH95LesHouches,Barenco95,Sleator95,DiVincenzo95,Domokos95}.  Cirac
and Zoller \Cite{Cirac95} have proposed the implementation of a QC
using laser cooled ions in a linear ion trap. The qubits are stored on
a metastable atomic transition, while the ions undergo collective
quantized motion in the linear trap \cite{Raizen92}, and interact
individually with laser light. Exchange of phonons stimulated by laser
pulses allows the implementation of quantum gates.  The distinctive
features of this system are: it allows the realization of quantum
gates between any set of (not necessarily neighboring) ions; in
comparison with other systems, decoherence is small, and the final
readout of the output state can be performed with essentially unit
efficiency employing the methods of quantum jumps. A first
experimental implementation for quantum gates along these lines has
been reported from the NIST Boulder group
\Cite{Wineland95}.
A QC based on optical CQED has been proposed by Pellizzari {\em et al.~}
\Cite{Pellizzari95}. It is based on a string of atoms trapped inside a high Q
optical cavity. Long lived
Zeeman coherences of ground state atoms represent the qubits.
Quantum gates between atoms are realized via exchange of a photon through the
single quantized cavity mode \cite{Pellizzari95}.  As in the case of
ion traps, quantum gates can be performed between any two qubits.
Sources of incoherence in this model are cavity decay and
spontaneous emission of atoms from the excited state during the gate
operation. However, spontaneous emission can be significantly avoided
by performing the gate operation between two atoms as an adiabatic
passage process via a dark state of the strongly coupled atom+cavity
system \cite{Parkins93}; and cavity decay is minimized by having a
photon present only during the gate operation.
\subsubsection{Errors and their correction}
There are two sources of errors in quantum computing. First, there are
{\em systematic errors}, for example, due to uncertainties in
adjustment of parameters. Second, there is {\em decoherence} as a
result of coupling to the environment
\Cite{Unruh95,Landauer95,Chuang95,Palma95}.  The general state of a register
of a QC is described by a density matrix
\be \Label{QCdm}
\rho \targ{t} =
\sum_{x,=0}^{2^L-1}\sum_{x'=0}^{2^L-1} \ket{x} \rho_{x,x'}(t)  \bra{x'} \; .
\ee
Thus as a result of the contact with the environment there will be
changes in populations (the diagonal density matrix elements), and
decay of the coherences (the off-diagonal density elements) on a time
scale determined by the coupling of the QC to the environment. The
power of quantum computing results from interference between quantum
paths, consequently a loss of coherence will destroy the quantum parallelism.
Unruh
\Cite{Unruh95} has shown that the computer--environment coupling leads
to an exponential error rate with the number of qubits $L$, and it has
been argued that this dependence will be the decisive factor to
limit the size of problems which can be solved on the QC. What these
limits are will be determined by fundamental limits of specific theoretical
proposals for the decay of  qubits and noise in
quantum gates, and by technological and experimental progress.
Specific models for dissipation have been studied in references
\Cite{Unruh95,Landauer95,Chuang95,Palma95}.

The above discussion of decoherence assumes that the environment is
traced out, i.e. in the language of continuous measurement theory ``the
results of the measurements in the out channels are not read.'' This
is the situation where the time evolution of the density matrix
(\Ref{QCdm}) is described by the master equation (a priori dynamics).
It is of interest to study the evolution of a QC which is continuously
monitored to detect the decay (quantum jump), and thus an incoherent
step in the evolution (compare nonlinear evolution Eq.~(\Ref{propwf},
and the quantum jump described by Eq.~(\Ref{qjwf}) \Cite{Pellizzari95}.
After detection of a decay the computer can be restarted; on the other
hand, by restricting the ensemble to those computer runs with no decay,
we improve the statistics for the read out. In the CQED example given
above this could be a cavity decay, or a photon from a spontaneous
emission of one of the atoms representing the qubits. Note, however,
that - even when {\em no decay is detected} in a specific run - {\em
information is gained} by this ``non-observation of a decay'' and this
leads to a distortion of the superposition of the qubits.  To illustrate
the effects of damping and continuous observation, we consider
the following simple example. Let us assume that we store a
superposition (qubit) $\ket{\psi (t\!=\!0)}=c_0 \ket{0} + c_1 \ket{1}$
on an atomic transition with ground state $\ket{0}$ and excited state
$\ket{1}$, where the upper state decays with a radiative rate
$\Gamma$.  At time $t\!=\!0$ we prepare the qubit in the superposition
$\ket{\psi (t\!=\!0)}=c_0 \ket{0} + c_1 \ket{1}$. If we do not monitor
the decay of the qubit ({\em a priori} dynamics), the time evolution
of the qubit is governed by the solution of the {\em master equation}:
\bea
\ket{\psi &(&t\!=\!0)} \bra{\psi (t\!=\!0)}
\rightarrow 
 \exp \Lc t \,   \ket{\psi (0)} \bra{\psi (0)}  \\
&=&
\abs{c_0}^2 \ket{0}\bra{0} \left( 1 - e^{-\Gamma t}  \right) +
\abs{c_1}^2 \ket{1}\bra{1}  e^{-\Gamma t} +
\left(c_1 c_0^*\ket{1}\bra{0} +  {\rm H.c.}   \right)  e^{-\half \Gamma t} \; .
\nn
\eea
We note that the initial pure state develops into a mixture, until the
atom finally has decayed to the ground state. If we monitor the system
continuously (with unit detection efficiency) and observe {\em no} decay ({\em
a
posteriori dynamics}), the state evolves according to (\Ref{propwf}) and
remains a pure state; however the
superposition is distorted by the decay (see also Ref.~\Cite{M93}):
\bea
\ket{\psi (t\!=\!0)}
&\rightarrow &
\ket{\psi (t)} = e^{ -i \Heff t} \ket{\psi (0)} / \norm{ e^{ -i \Heff t} \ket{
\psi (0)}  } \nn %
\\
&=&
\nofrac{
\left( c_0 \ket{0} + c_1 \ket{1} e^{- \half \Gamma t} \right)
}{\norm{
\ldots
}} %
\\
&\rightarrow& \ket{0} \quad \mbox{for $t \rightarrow \infty$} \; . \nn
\eea
This unwanted dynamics of the qubits corresponds to our increase of
information which we gain by observing (measuring) the system: by not
registering a decay our knowledge increases that the system is in the
ground state. The distortion of the qubit is a result of the asymmetry of the
decay. When we flip the qubit at time $t/2$, $\ket{0} \leftrightarrow \ket{1}$ , and again at time $t$, we symmetrize  the decay,
\bea
\ket{\psi (t\!=\!0)}
&\rightarrow&
\nofrac{
\left( c_0 \ket{0} e^{- \frac{1}{4} \Gamma t} + c_1 \ket{1} e^{- \frac{1}{4}
\Gamma t} \right)
}{\norm{
\ldots
}} %
\\
&=& c_0 \ket{0}  + c_1 \ket{1} \equiv \ket{\psi (t\!=\!0)} \; . \nn
\eea
Thus, the information which state has decayed has been lost, the
superposition remains unchanged, and there is no effect of decay (in
the subensemble with no decay).  The idea of a symmetric decay, which
factors out in the time evolution (\Ref{propwf}) and thus drops out
upon normalization, is readily generalized to a {\em construction of
gates}. This problem is studied in more detail in
Ref.~\Cite{Pellizzari95}. It is an interesting question to what extent
schemes can be found where the state of the register can be {\em
reconstructed} after a quantum jump \Cite{Mabuchi95}.

\vspace{1cm} Acknowledgment: We thank Klaus Ellinger, Simon Gardiner,
Klaus Gheri, Thomas Pellizzari and Herwig Stecher for reading the
manuscript.  P.Z. thanks F. Haake for an exciting climb of Mont Blanc
during the summer school.

\appendix

\section{An exercise in Ito calculus}
\Label{ItoExcercise}

As an exercise in \Ito\/ calculus we will go in some detail through
the steps of deriving Eq.(\Ref{SSEn}) from (\Ref{SSEu}). To simplify
notation we confine ourselves to the case of a single channel $\Nd=1$.
We will show that if $\pct (t)$ obeys
\be \Label{ex1}
d \pct(t) = [ A \: dt + (B -\one ) \dN{}{t} ]
\pct (t)
\ee
with $A$ and $B$ operators, then the normalized $\psi_c (t)$ obeys
\be \Label{ex2}
d \psi_c(t) = [ (A-\AVc{A+A^\dagger}) \: dt + (B / \sqrt{\AVc{B^\dagger
B}} - 1 ) \dN{}{t} ]
\psi_c (t)
\ee
with $\AVc{\ldots}$ defined in Eq.~(\Ref{expect}).

Proof: From
\be
\pct(t\!+\!dt) = [\one + A\: dt + (B-\one) \dN{}{t}] \pct(t)
\ee
the change of the norm in the time step $dt$ is
\be
\norm{\pct (t\!+\!dt)}^2 = \norm{\pct(t)}^2
\left[1 + \AVc{A+A^\dagger} \: dt + \AVc{B^\dagger B-\one} \dN{}{t}\right]
\ee
and
\bea
\norm{\pct (t\!+\!dt)}^{-1} &=& \norm{\pct (t)}^{-1}
[1 + \AVc{A+A^\dagger} \: dt + \AVc{(B^\dagger B-\one)} \dN{}{t}]^{-1/2}  \nn%
\\
\nn &=&
\norm{\pct (t)}^{-1}
[1 - \half \AVc{A+A^\dagger} \: dt + (1/\sqrt{\AVc{B^\dagger B}}-1)\;
\dN{}{t}] \; .
\eea

This follows from
\bea
f(1+a \: dt + b\; dN)&=& f(1)+f'(1)\; a \: dt +f'(1)\; b \; dN +
\frac{1}{2!} f''(1)\; b^2 \; dN^2 + \ldots \nn%
\\  \nn
&=& f(1) +f'(1)\; a\: dt + [f(1+b) - f(1)]\; dN
\eea
valid for sufficiently well-behaved functions $f$.
Finally, we have
\bea
\psi_c (t\!+\!dt) &=& [\one + A\: dt + (B-\one) \dN{}{t}] \psi_c(t) \times %
\\
 && [1 - \half \AVc{A+A^\dagger} \: dt + (1/\sqrt{\AVc{B^\dagger B}}-1)\;
\dN{}{t}] \nn
\eea
which gives (\Ref{ex2}).




\begin{thebibliography}{999}


\Bibitem{QN} C.W. Gardiner,
    {\em Quantum Noise} (Springer, Berlin, 1991).


\Bibitem{Carmichael} H.J.Carmichael,
{\em An Open Systems Approach to Quantum Optics}, Lectures Notes in Physics m18
(Springer, Berlin, 1993)

\Bibitem{WallsMilburn} D.F. Walls and G. Milburn, {\em Quantum Optics}
(Springer, Berlin, 1995).

\Bibitem{Louisell} W. Louisell, {\em Quantum Statistical Theories of Radiation}
(Wiley, New York, 1974, 1989).

\Bibitem{HP} R. Hudson and K.R. Parthasarthy, Commun.
         Math. Phys. {\bf 93}, 301 (1984).


\Bibitem{SM} C.W. Gardiner, {\em Handbook of Stochastic
    Methods}, 2nd Ed.~(Springer, Berlin 1985, 1989).


\Bibitem{BL85} A. Barchielli and G. Luperi,  J. Math. Phys. {\bf 26}, 1985
(1985).
\Bibitem{B86} A. Barchielli, Phys. Rev. A {\bf 34}, 1642 (1986)
\Bibitem{B87} A. Barchielli, J. Phys. A {\bf 20}, 6341 (1987).
\Bibitem{B90}  A. Barchielli, Quantum Opt. {\bf 2}, 423 (1990)
\bibitem{BB91}  A. Barchielli and V. P. Belavkin, J. Phys. A {\bf 24}, 1495
(1991) and references cited.

\Bibitem{GardinerCollett} C.W. Gardiner and M.J.
         Collett, Phys. Rev. A {\bf 31}, 3761 (1985).


\Bibitem{GPZ92} C. W. Gardiner, A. S. Parkins, and P. Zoller, Phys. Rev. A {\bf
46}, 4363 (1992).


\Bibitem{WisemanThesis} H. M. Wiseman, {\em Quantum Trajectories and Feedback},
PhD Thesis, University of Queensland, June 1994;
 H. M. Wiseman and G. Milburn, Phys. Rev. A {\bf 47}, 642 (1993); {\em ibid.}
1652 (1993).


\Bibitem{SD81} M.D. Srinivas and E.B. Davies,
    Optica Acta {\bf 28}, 981 (1981).

\Bibitem{G94} P. Goetsch and R. Graham, Phys. Rev. A {\bf 50}, 5242 (1994);  P.
Goetsch, R. Graham and F. Haake, Phys. Rev. A {\bf 51} 136 (1995).




\Bibitem{DCM92} J. Dalibard, Y. Castin, and K. M{\o}lmer, Phys. Rev.
Lett.  {\bf 68}, 580 (1992).

\Bibitem{M93} K. M{\o}lmer, Y. Castin and J. Dalibard,  J. Opt. Soc. Am. B, {
\bf 10}, 524 (1993).

\Bibitem{DRZ92} R. Dum, P. Zoller, and H. Ritsch, Phys. Rev. A
{\bf 45}, 4879 (1992).

 \Bibitem{DPZG92} R. Dum, A.S. Parkins, P. Zoller, and C.W. Gardiner,
Phys. Rev. A {\bf 45}, 4879 (1992).

\Bibitem{TC92} L. Tian and H. Carmichael, Phys. Rev. A {\bf 46}, R6801 (1992).

\Bibitem{N95} M Naraschewski and A. Schenzle, Z. Phys. D {\bf 33}, 79 (1995)

\Bibitem{Bel} V. P. Belavkin, J. Phys. A. {\bf 22}, L1109 (1989);
 V. P. Belavkin, J. Math. Phys. {\bf 31}, 2930 (1990);
 V. P. Belavkin and P. Staszewski, Phys. Rev. A {\bf 45}, 1347 (1992)

\Bibitem{CBA92} C. Cohen-Tannoudji, F. Bardou, and A. Aspect, in
{\it Laser Spectroscopy X},
edited by M. Ducloy, E. Giacobino, and G. Camy (World
Scientific, 1992).


\Bibitem{Marte}
P. Marte, R. Dum, R. Taieb, P. D. Lett and P. Zoller, \Title{Quantum
wave function
simulation of the resonance fluorescence spectrum from
one-dimensional optical
molasses} Phys. Rev. Lett. {\bf 71}, 1335 (1993);
 P. Marte, R. Dum, R. Taieb and P. Zoller,
\Title{Resonance fluorescence from quantized one-dimensional
molasses}
Phys. Rev. A {\bf 47}, 1378 (1993).

\Bibitem{Dum94}
R. Dum, P. Marte, T. Pellizzari and  P. Zoller ,
\Title{Laser cooling to a single quantum state  in a trap},
Phys Rev. Lett. {\bf 73}, 523 (1994).

\Bibitem{CBSDM94} Y. Castin, K. Berg-Sorensen, J. Dalibard, K. Molmer,
Phys. Rev. A {\bf 50}, 5092 (1994)


\Bibitem{NKS95} G. Nienhuis, J. de Kloe, and  P. van der Straten,
JOSA B {\bf 12}, 520 (1995)



\Bibitem{CM95} Y. Castin, and  K. Molmer, Phys. Rev. Lett. {\bf 74}, 3772
(1995)


\Bibitem{SGK95} J. Steinbach, B. M. Garraway, and P. L. Knight,
Phys. Rev. A {\bf 50}, 3302 (1995)



\Bibitem{GK94a}B. M. Garraway, and P. L. Knight, Phys. Rev. A {\bf 50}, 2548
(1994)

\Bibitem{GK94b}B. M. Garraway, and P. L. Knight, Phys. Rev. A {\bf 49}, 1266
(1994)




\Bibitem{C93} H. J. Carmichael, Phys. Rev. Lett. {\bf 70}, 2273 (1993)
;
 C. W. Gardiner, Phys. Rev. Lett. {\bf 70}, 2269 (1993)

\Bibitem{G84} N. Gisin, Phys. Rev. Lett. {\bf 52}, 1657
(1984); N. Gisin, Helv. Phys. Acta {\bf 62}, 363 (1989).
\Bibitem{GP92} N. Gisin and I.C. Percival, Phys. Lett. A {\bf 167}, 315 (1992);
 J. Phys. A {\bf 25}, 5677 (1992); J. Phys. A {\bf 26}, 2233 (1993); N. Gisin,
P. L. Knight, I. C. Percival, R. C. Thompson and D. C. Wilson, J. Mod. Opt. {
\bf 40}, 1663 (1993); N. Gisin,J. Mod. Opt. {\bf 40}, 2313 (1993); Y. Salama
and N. Gisin, Phys. Lett. A {\bf 181}, 269 (1993).

\Bibitem{Stenholm} S. Stenholm, Physica Scripta {\bf 47}, 724 (1993).

\Bibitem{DGHP95} L. Diosi, N. Gisin, J. Halliwell, and I. C. Percival,
 Phys. Rev. Lett. {\bf 74}, 203 (1995)



\Bibitem{ZMW87} P. Zoller, M. Marte and D. F. Walls, Phys. Rev. A {\bf 35}
(1987);
 R. Blatt, W. Ertmer, J. Hall and P. Zoller,
 Phys. Rev. A {\bf 34}, 3022 (1986).


\Bibitem{M75} B.R. Mollow, Phys. Rev. A {\bf 12}, 1919 (1975).


\Bibitem{BZ} R. Blatt and P. Zoller, Europ. J. Phys. {\bf 9}, 250
(1988); R. J. Cook,  Prog. Opt., {\bf 28}, 361 (1990).


\Bibitem{qjexp}
 W. Nagourney, J. Sandberg, H. Dehmelt, Phys. Rev.
Lett. {\bf 56}, 2797 (1986); J. C. Bergquist, Randall G. Hulet, Wayne M.
Itano, and D. J. Wineland, Phys. Rev. Lett. {\bf 57}, 1699 (1986); Th.
Sauter, W. Neuhauser, R. Blatt, and P. E. Toschek, Phys. Rev. Lett. {\bf 57
}1696 (1986).

\Bibitem{Holland95} M. Holland, S. Marksteiner, P. Marte and P. Zoller,
Phys. Rev. Lett. {\bf 76} 3683 (1996).

\Bibitem{K95LesHouches} See lecture notes by P. Knight, this volume.

\Bibitem{Walther94} H. Walther, {\em Adv. in At. Mol. and Opt. Phys.}, {\bf
32}, 379 (1994).

\bibitem{Blatt95}  R. Blatt, Proc. 14-th ICAP, ed. S. Smith, C. Wieman, and D.
Wineland
{\em et al.~} (AIP Press, 1995), p. 219.

\Bibitem{CD86} C. Cohen--Tannoudji and  J. Dalibard, Europhys. Lett.
{\bf 1}, 441 (1986).



\Bibitem{Cirac93}
J. I. Cirac, R. Blatt, A. S. Parkins and P. Zoller,
\Title{Preparation of Fock states by observation of quantum jumps in
an ion trap}
Phys. Rev. Lett. {\bf 70}, 762 (1993);  J. I. Cirac, A. S. Parkins, R. Blatt
and P. Zoller,
Phys. Rev. Lett. {\bf 70}, 556 (1993).


\Bibitem{Toschek95} J. Eschner, B. Appasamy and P. E. Toschek, Phys. Rev. Lett.
{
\bf 74}, 2435 (1995).
\Bibitem{RH95LesHouches} See lecture notes on CQED by J. M. Raimond, and S.
Haroche, this volume.




\Bibitem{MolmerCastin95} K. M\o lmer and Y. Castin, 
Quantum and Semiclass. Opt. {\bf 8} 49 (1996) .

  \Bibitem{Jessen} P.S. Jessen, C. Gerz, P. D. Lett, W. D. Phillips, S. L.
Rolston, R. J. C. Spreeuw, C. I. Westbrook,  Phys.  Rev. Lett.
{\bf
69}, 49 (1992).
 \Bibitem{Verkerk} P. Verkerk, B. Lounis, C. Salomon, C. Cohen-Tannoudji, J.
Y.
Courtois and G. Grynberg,  Phys. Rev. Lett.
{\bf
68}, 3861 (1992);   G. Grynberg  B. Lounis, P. Verkerk, J.-Y. Courtois, C.
Salomon, Phys. Rev. Lett.
{\bf 70}, 2249 (1993).
\Bibitem{HemmerichHansch} A. Hemmerich and T. W. H\"ansch,
Phys. Rev.  Lett. {\bf 70}, 410 (1993).




\Bibitem{Wallis} H. Wallis, Physics Reports {\bf 255}, 203 (1995).
 \Bibitem{CastinDalibard} Y. Castin and J. Dalibard, Europhys. Lett.
{\bf 14}, 761 (1991).
 \Bibitem{DalibardCohen} J. Dalibard and C.  Cohen--Tannoudji, J. Opt.
Soc.  Am. B {\bf 6}, 2023 (1989); J. Dalibard and C. Cohen-Tannoudji,
J. Opt. Soc.  Am. B {\bf 2}, 1707 (1985).

\Bibitem{Pfau}  T. Pfau, S. Spalter, Ch. Kurtsiefer, C. R. Ekstrom, J. Mlynek, Phys. Rev. Lett.
                {\bf 73}, 1223 (1994).
\Bibitem{radiativeHeating} M. J. Holland , K.-A. Suominen, and K. Burnett,
                Phys. Rev. Lett. {\bf 72}, 2367 (1994).




\bibitem{Ekert95}  For an review see, A. Ekert, Proc. 14-th
ICAP, ed. S. Smith, C. Wieman and D. Wineland (AIP Press, 1995) p. 450.; R.
Josza and A. Ekert, to appear in Rev. Mod. Phys.

\bibitem{Shor94}  P. Shor, Proc. 35-th Annual Symposium on
Foundations of Computer Science, IEEE Press (1994).

\bibitem{Unruh95}   W. G.
Unruh, Phys. Rev. A {\bf 51}, 992 (1995).

\bibitem{Landauer95} R. Landauer, Proc. Royal Soc. London A {\bf 353} 367
(1995).

\Bibitem{Chuang95}  I. Chuang, R. Laflamme, P.
Shor, and W. Zurek, Science {\bf 270} 1633 (1995).

\Bibitem{Palma95}  G. M. Palma, K.A. Suominen, and A. Ekert,
Proc. Royal Soc. London A {\bf 452} 567 (1996).

\bibitem{Barenco95}  A. Barenco, D. Deutsch, and A. Ekert, Phys. Rev. Lett.
{\bf 74}, 4083 (1995).

\Bibitem{Sleator95}  T. Sleator and H. Weinfurter, Phys. Rev. Lett.
{\bf 74}, 4087 (1995).

\Bibitem{DiVincenzo95} D.~P.~DiVincenzo Phys. Rev. A {\bf 51}, 1015
(1995)

\Bibitem{Domokos95} P. Domokos, J. M. Raimond, M. Brune, and S. Haroche,
Phys. Rev. A {\bf 52} 3554 (1995).

\bibitem{Cirac95}  J. I. Cirac and P. Zoller, Phys. Rev. Lett. {\bf 74},
4091 (1995).

\bibitem{Raizen92}  M. G. Raizen, J. M. Gilligan, J. C. Bergquist, W. M.
Itano, and D. J.  Wineland, Phys. Rev. A {\bf 45}, 6493 (1992).

\bibitem{Wineland95}  C. Monroe, D. M. Meekhof, B. E. King, W. M. Itano,
and D. Wineland, Phys. Rev. Lett. {\bf 75} 4714 (1995).


\bibitem{Pellizzari95}  T. Pellizzari, S. A. Gardiner, J. I. Cirac, and P.
Zoller,  Phys. Rev. Lett. {\bf 75} 3788 (1995).





\bibitem{Parkins93} A. S. Parkins, P. Marte, P. Zoller, and H. J. Kimble, Phys.
Rev. Lett. {\bf 71}, 3095 (1993).


\Bibitem{Mabuchi95} P.Z. thanks A. Barenco and H. Mabuchi for discussion of these
ideas during the school.

\end{thebibliography}
\end{document}